\documentclass[sigconf]{acmart}
\AtBeginDocument{%
  }
\usepackage{makecell}
\usepackage{multirow} 
\usepackage{colortbl} 
\usepackage{color}
\definecolor{lightpinkt}{rgb}{0.9, 0.8, 0.8}
\definecolor{darkpinkt}{rgb}{0.9, 0.65, 0.65} 
\definecolor{deeppinkt}{rgb}{0.9, 0.5, 0.5}
\acmConference[CHI '25]{Proceedings of the CHI Conference on Human Factors in Computing Systems}{Apr 26–May 1}{Yokohama, Japan}
\acmBooktitle{Proceedings of the CHI Conference on Human Factors in Computing Systems (CHI '25), Apr 26–May 1, 2025, Yokohama, Japan}

\acmISBN{978-1-4503-XXXX-X/18/06}

\begin{document}


\title{Since U Been Gone: Augmenting Context-Aware Transcriptions for Re-engaging in Immersive VR Meetings}




\settopmatter{authorsperrow=4}
\author{Geonsun Lee}
\orcid{0000-0001-9401-8559}
\affiliation{%
  \institution{University of Maryland}
  \city{College Park}
  \state{MD}
  \country{USA}
}
\email{gsunlee@umd.edu}

\author{Yue Yang}
\orcid{0000-0001-7047-9801}
\affiliation{%
  \institution{Stanford University}
  \city{Stanford}
  \state{CA}
  \country{USA}
}
\email{yyang198@stanford.edu}

\author{Jennifer Healey}
\orcid{0000-0002-5700-4921}
\affiliation{%
  \institution{Adobe Research}
  \city{San Jose}
  \state{California}
  \country{USA}
}
\email{jehealey@adobe.com}

\author{Dinesh Manocha}
\orcid{0000-0001-7047-9801}
\affiliation{%
  \institution{University of Maryland}
  \city{College Park}
  \state{MD}
  \country{USA}
 }
\email{dmanocha@umd.edu}

\renewcommand{\shortauthors}{Lee et al.}

\begin{abstract}

Maintaining engagement in immersive meetings is challenging, particularly when users must catch up on missed content after disruptions. While transcription interfaces can help, table-fixed panels have the potential to distract users from the group, diminishing social presence, while avatar-fixed captions fail to provide past context. We present EngageSync, a context-aware avatar-fixed transcription interface that adapts based on user engagement, offering live transcriptions and LLM-generated summaries to enhance catching up while preserving social presence. We implemented a live VR meeting setup for a 12-participant formative study and elicited design considerations. In two user studies with small (3 avatars) and mid-sized (7 avatars) groups, EngageSync significantly improved social presence ($p < .05$) and time spent gazing at others in the group instead of the interface over table-fixed panels. Also, it reduced re-engagement time and increased information recall ($p < .05$) over avatar-fixed interfaces, with stronger effects in mid-sized groups ($p < .01$).
\end{abstract}

\begin{teaserfigure}
  \includegraphics[width=\textwidth]{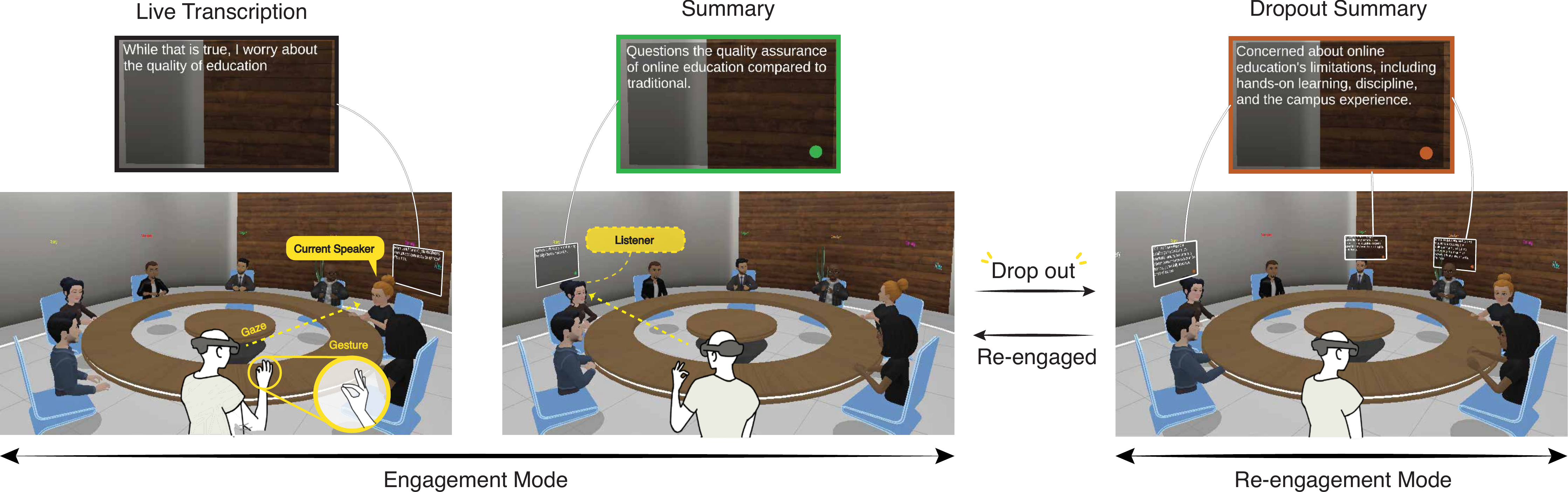}
  \caption{\textbf{EngageSync} is a context-aware transcription panel designed to help users in immersive VR meetings catch up on conversations while maintaining social presence. It operates in two modes: \textbf{Engagement Mode} and \textbf{Re-engagement Mode}. In Engagement Mode, users can view live transcriptions of the speaker or a summary of the previous utterance of a listener by gazing at the person of interest and performing a pinch gesture. In Re-engagement Mode, summaries of conversations that occurred during the user’s absence are displayed for all relevant participants. Once all summaries are read, the user is considered "re-engaged" and caught up with missed context, and the system returns to Engagement Mode.}
  \Description{A visual comparison of EngageSync's two operational modes: Engagement Mode and Re-engagement Mode. In Engagement Mode, a user gazes at avatars and performs a pinch gesture to view live transcriptions of the current speaker (default text panel) or a summary of a listener's previous utterance (text panel with a green circle on the bottom right). In Re-engagement Mode, summaries of conversations that occurred during the user's absence are displayed for all relevant participants (text panel with an orange circle on the bottom right). Once all summaries are read, the user is considered "re-engaged" and the system returns to Engagement Mode. The figure illustrates how EngageSync adapts based on user engagement, helping maintain social presence while providing efficient ways to catch up on missed content.}
  \label{fig:teaser}
\end{teaserfigure}

\received{20 February 2007}
\received[revised]{12 March 2009}
\received[accepted]{5 June 2009}

\maketitle

\section{Introduction}
Remote work and virtual collaboration have become integral to modern communication with video-mediated meetings seeing unprecedented growth following the COVID-19 pandemic~\cite{karl2022virtual,lee2022distracting}.


As the demand for seamless remote interaction continues to grow, researchers and developers have turned their attention not only to video conferencing platforms but also to more immersive technologies like Virtual Reality (VR) and Mixed Reality (MR)~\cite{mcveigh2022beyond}. These technologies offer new opportunities to reimagine how we engage in virtual environments (VEs), providing enhanced social interaction and presence that video conferencing alone cannot achieve.~\cite{mcveigh2021case, mcveigh2022beyond,park2024lessons}

While remote meetings offer significant convenience, they also make users more susceptible to distractions, both internal (e.g., multitasking or loss of focus) and external (e.g., notifications, people entering the room)~\cite{cao2021large}. Research in remote video meeting interfaces has made strides in keeping users engaged through meeting summaries, post-meeting transcripts~\cite{samrose2021meetingcoach}, and more recently, live interactive transcription tools designed to maintain participation during discussions~\cite{son2023okay,chen2023meetscript}. These tools not only help users stay on track but also enable them to catch up after disruptions, which can occur frequently in remote work settings.

Although immersive VR platforms promise heightened focus and social presence through their immersive nature, users in VR environments are not immune to external distractions. Interruptions such as device notifications, someone physically entering the room, or technical difficulties can pull users out of their virtual space. Previous work has attempted to address these challenges in VR by mitigating disruptions during experiences~\cite{gottsacker2021diegetic, ghosh2018notifivr}. 
However, questions arise about whether video meeting transcription interfaces can be directly applied to the spatial, immersive nature of VR meetings.

One possible solution is to replicate video meeting interfaces by providing a spatially fixed transcription panel in front of the user, allowing them to scroll through transcripts. However, this approach often forces users to look away from other meeting participants, breaking social engagement and reducing the sense of presence in the group. Current social VR platforms, such as VRChat~\cite{VRChat}, handle this issue by attaching captions directly to the speaking avatar, maintaining eye contact and social interaction. Yet, these avatar-fixed captions, do not provide missed content, particularly when users return from a disruption and need to catch up.

This raises the question of how we can bring the advantages of transcription interfaces from video meetings into immersive VR environments while balancing the need for both social presence and effective re-engagement after interruptions.
To address this question, we first conducted a formative study examining how users interact with live transcriptions in immersive VR meetings when experiencing disruptions. 
Using a conventional table-fixed interface, we observed how participants dropped out and rejoined conversations, analyzing their catch-up strategies and social connection maintenance.
This investigation revealed key challenges in designing transcription interfaces that support both efficient re-engagement and social presence in VR environments. Based on these insights, we developed \textbf{EngageSync}, a novel, context-aware avatar-fixed transcription interface that dynamically adapts to the user’s engagement state. EngageSync supports two key use cases: (1) providing real-time transcription and summaries during active participation, and (2) offering context-aware summaries to support re-engagement after disruptions. By adapting based on whether users are engaged or re-engaging, EngageSync maintains social presence while ensuring efficient catch-up on missed content.

We evaluate EngageSync through two user studies, comparing it to existing interfaces such as table-fixed and always-on avatar-fixed transcription panels. Our evaluation investigates its effectiveness in enhancing social presence and information recall across different group sizes and scenarios.

This work makes the following contributions:

\begin{itemize} 
    \item Design insights for context-aware transcription in immersive VR environments: Based on a formative study with 12 participants, we identify key design challenges for supporting re-engagement and enhancing social presence in immersive group discussions.
    
    \item EngageSync, a novel adaptive avatar-fixed transcription interface: EngageSync provides live transcriptions for active conversations and context-aware summaries for re-engagement supporting users to balance social presence with effective conversation catch-up.

    \item Comprehensive evaluation across group sizes: Through two user studies, we evaluate EngageSync’s performance compared to table-fixed and avatar-fixed interfaces, demonstrating significant improvements in social presence, information recall, and faster re-engagement times ($p<.05$).
    
    \item Exploration of group size effects on interface performance: Our results show how EngageSync’s context-aware features become more beneficial in larger groups, providing important insights for the design of future immersive meeting 
    platforms.
\end{itemize}

\section{Related Works}
\subsection{Meeting Engagement, Distractions, and Eye-contact}

\subsubsection{Meeting Engagement and Participation} 


Maintaining engagement in virtual meetings is challenging due to the loss of non-verbal cues like eye contact and subtle gestures, which are essential for signaling attentiveness and turn-taking~\cite{schwartzman1989meeting}. This lack of feedback can lead to misinterpretations of engagement levels, making it harder for participants to contribute effectively.

To address this, Hosseinkashi et al.~\cite{hosseinkashi2024meeting} developed a system that detects failed interruptions, helping ensure that overlooked participants have a chance to speak. Beyond speech cues, Oertel et al.~\cite{oertel2013gaze} analyzed gaze and posture to assess group engagement, showing that sustained mutual gaze and synchronized gestures indicate higher participation. 

These insights show the importance of designing systems that help users re-engage naturally after disruptions. Our approach builds on this by integrating context-aware transcripts that restore conversational flow while preserving engagement within immersive environments.

\subsubsection{Distractions in Remote Meetings and VR} 
Maintaining focus is a known challenge in remote meetings, with multitasking occurring in roughly 30\% of meetings, significantly reducing engagement~\cite{vedernikov2024analyzing, cao2021large}. Common distractions include personal obligations, household chores, and non-meeting-related activities like checking emails~\cite{karl2022virtual, lee2022distracting, cao2021large}. Technical issues such as connectivity problems and background noise further exacerbate these distractions~\cite{guetter2022person}. Eye-tracking studies show participants often look away from their screens, particularly in smaller groups~\cite{george2022users}.

Although VR reduces visual distractions from the physical world, external interruptions like notifications or environmental sounds can disrupt users~\cite{gottsacker2021diegetic, rzayev2019notification, ghosh2018notifivr}. While VR environments enhance spatial presence~\cite{park2024lessons}, external disruptions become more jarring and break immersion~\cite{gottsacker2021diegetic}. Unlike diegetic solutions for managing distractions~\cite{gottsacker2021diegetic}, our focus is on helping users re-engage after disruptions.

\subsubsection{Attention and Social Presence in Remote Communication} 

To foster trust and facilitate collaboration in meetings, being aware of others' focus is crucial~\cite{hietanen2018affective, kaiser2022eye}. Hence it is important to mitigate the absence of gaze cues in remote communication to preserve the sense of social presence. He et al. developed GazeChat, which visually represents users as gaze-aware 3D profile photos using neural rendering, helping users share their direction of attention~\cite{he2021gazechat}. In a collaborative immersive VR environment, this sense of social presence becomes more evident~\cite{lee2020user}. EyeCVE uses mobile eye-trackers
to drive the gaze of each participant’s virtual avatar, thus supporting remote mutual eye contact and awareness of others’ gaze in a CAVE system~\cite{steptoe2008eye}. Other research has explored the impact of explicitly visualizing users’ gaze in task-oriented collaboration~\cite{amores2015showme, piumsomboon2019effects}.

In our work, we do not explore explicit gaze visualization, which can be distracting or intrusive in conversational contexts, and instead use gesture-triggered avatar-fixed panels that direct attention toward the other speakers.

\subsection{Transcription Interfaces in Video Meetings}

Maintaining engagement and focus in remote meetings is a well-known challenge. While much of the attention has been on fostering awareness of participants' attention and mutual presence, another line of research focuses on how transcription interfaces can help users stay engaged with meeting content, particularly in situations where they may become distracted or miss parts of the conversation.

\subsubsection{Transcription Interfaces}
Transcription interfaces have evolved significantly over the years to support participants in engaging with and reflecting on meeting content. Whittaker et al. demonstrated that reading a transcript is often more efficient than rewatching video or audio recordings of meetings, as transcripts allow users to quickly skim and locate relevant information~\cite{whittaker2008design}. Moreover, studies have shown that enabling interaction with transcripts, such as marking or highlighting keywords, can facilitate information recall and reduce users' cognitive load~\cite{kafle2019evaluating}.

Post-meeting transcription tools has evolved, focusing on generating summaries and providing feedback after meetings have ended. For example, Banerjee et al. explored methods for generating summaries post-meeting to help users efficiently digest key takeaways~\cite{banerjee2015generating}. Moreover, the MeetingCoach tool supports users to reflect on past meetings by providing personalized feedback on their performance and participation~\cite{samrose2021meetingcoach}. 

\subsubsection{Real-time Interactive Transcription Interfaces}
In recent years, the focus has also shed light on real-time interactive transcription tools that keep participants engaged during the meeting itself. For instance, Zhang et al. found that generating summaries of chat-based meetings in real-time helps participants quickly catch up on missed discussions and enhances overall engagement~\cite{zhang2018making}. Tucker et al. developed and evaluated a "Catchup audio player" designed specifically for participants who join meetings late. The system automatically identifies the gist of what was missed, allowing latecomers to quickly catch up and participate effectively without needing to process the full transcript~\cite{tucker2010catchup}.

MeetScript is another example of a real-time transcription interface that allows participants to actively interact with the transcript as the conversation unfolds, marking significant moments and revisiting key sections in real-time~\cite{chen2023meetscript}. This approach shifts the focus from post-meeting reflection to active participation. Similarly, Iijima et al. introduced interactive text clouds to assist Deaf and Hard of Hearing (DHH) users in staying engaged with live video conferences~\cite{iijima2021word}. Son et al. introduced OPARTs~\cite{son2023okay}, a meeting interface that allows users to seamlessly switch between live transcription, summary modes, and keyword extracts from each speaker’s utterance to support users catching up with missed context due to internal and external distractions.

Although these advancements have been made in video-mediated meetings, adapting interactive transcription and
summarization tools to immersive environments like VR presents new challenges and opportunities which we will
address more in detail in the section below.

\subsection{Transcription and Engagement in Immersive VR Meetings}

Immersive VR meetings, often viewed as part of the broader metaverse, are emerging as a powerful tool for remote collaboration. These platforms offer opportunities to replicate and enhance traditional office interactions, with features such as spatial audio, natural gestures, and shared virtual spaces designed to improve communication and social presence~\cite{park2024lessons, mcveigh2022beyond}. As remote and hybrid work models continue to evolve, immersive VR meetings are increasingly seen as a solution for creating engaging, collaborative workspaces that foster both social presence and productivity~\cite{steinicke2020first}.

Several VR meeting platforms, including Mozilla Hubs~\cite{MozillaHubs}, Spatial~\cite{SpatialVR}, and Meta Horizon Workrooms~\cite{MetaHorizonWorkrooms}, have gained popularity by providing immersive environments where participants can interact in ways that closely simulate face-to-face meetings. These platforms report increased engagement and interaction due to the immersive nature of VR, but they also present unique challenges that need to be addressed for effective user experiences~\cite{abramczuk2023meet}.

\begin{figure*}[h]
	\centering
    \includegraphics[width=0.9\textwidth]{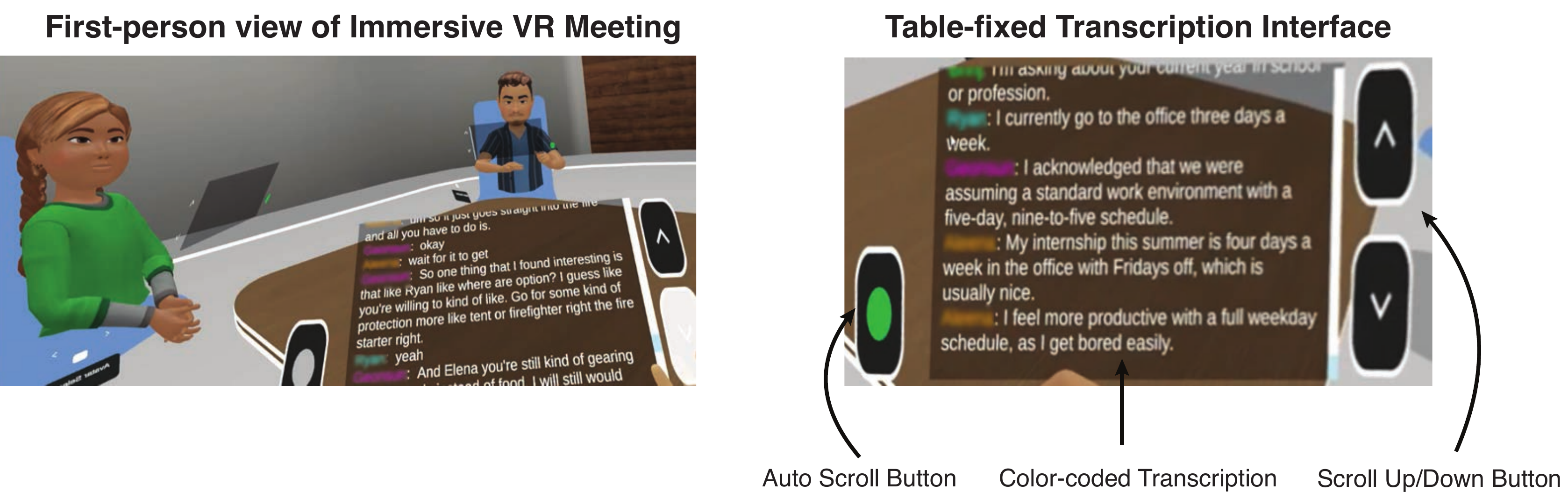}
	\vspace{-1.5em}
	\caption{The formative study setup of a four-people meeting. A screenshot of the immersive meeting environment from a first-person view (Left). The text panel interface was used in the formative study. The interface consists of a text panel where the participants' names are color-coded for readability, an auto-scroll button to the left that follows the newest lines in the panel, and a scroll-up and down button to the right (Right).}
    
\label{fig:formative_setup}
\end{figure*}

\subsubsection{Challenges in Immersive VR Meetings} Immersive VR environments, while offering many advantages for remote work, come with technical and social challenges. Issues such as hardware limitations, network latency, and avatar fidelity can hinder the overall user experience~\cite{park2024lessons}. Furthermore, users must navigate new social norms and behaviors in these VEs, which can impact professional interactions~\cite{dincelli2022immersive}.

To overcome these challenges, researchers have proposed various solutions. For example, Qian et al. introduced ChatDirector, which converts RGB video streams into 3D portrait avatars, overcoming the need for VR-specific equipment while enhancing the realism of VR meetings~\cite{qian2024chatdirector}. Similarly, ViGather bridges the gap between traditional devices (laptops, desktops) and VR environments, reconstructing users’ poses and conveying non-verbal cues like eye contact to enhance social presence~\cite{qiu2023vigather}. To overcome the limited field of view of VR headsets, Lee et al., implemented a multi-modal attention guidance system that leverages light and spatial audio so users can notice new speakers in a VR meeting~\cite{lee2024may}.

\subsubsection{Maintaining Engagement through Non-verbal Cues} 
Non-verbal cues like eye contact and spatial behavior are crucial for maintaining social presence in VR. Wang et al.~\cite{wang2024socially} found that proxemics and mutual gaze awareness enhance social connection and perceived attention during group interactions.

However, visualizing gaze or non-verbal cues in VR can sometimes distract users and pull them away from the conversation~\cite{herrera2020effect, kyrlitsias2022social}. To avoid this, our approach focuses on promoting natural engagement with avatars without explicitly visualizing gaze, ensuring the flow of conversation remains uninterrupted.

\subsubsection{Transcription, Captions, and Accessibility in AR/VR} Live transcription and captions have been widely used in video conferencing and Augmented Reality (AR) to improve accessibility and comprehension. Systems like Wearable Subtitles~\cite{olwal2020wearable} provide a comprehensive solution for supporting DHH users, combining hardware AR glasses with continuous speech transcription, translation, and sound awareness. Jain et al. conducted a preliminary study using HoloLens to prototype AR caption placement and collect contextual insights to inform future design considerations for AR-based accessibility tools~\cite{jain2018exploring}. In gaze-assisted interfaces such as StARe~\cite{rivu2020stare} reveal relevant information progressively during conversations, allowing for smoother interactions. These systems have proven valuable for enhancing accessibility, but they also introduce challenges in immersive settings where prolonged focus on captions can detract from the overall sense of presence~\cite{ubur2024easycaption}.

In social VR platforms like VRChat, captions are often used to enhance communication. However, captions can disrupt immersion if they dominate the user’s visual attention for too long. Uber et al. found that Automatic speech recognition (ASR) captions improve accessibility but can lead to disengagement from the conversation and reduce social connection if exposure time is prolonged~\cite{ubur2024easycaption}. Thus, there is a need for careful consideration in designing immersive transcription interfaces that maintain a balance between accessibility and engagement.

\begin{figure*}[h]
\centering\includegraphics[width=\textwidth]{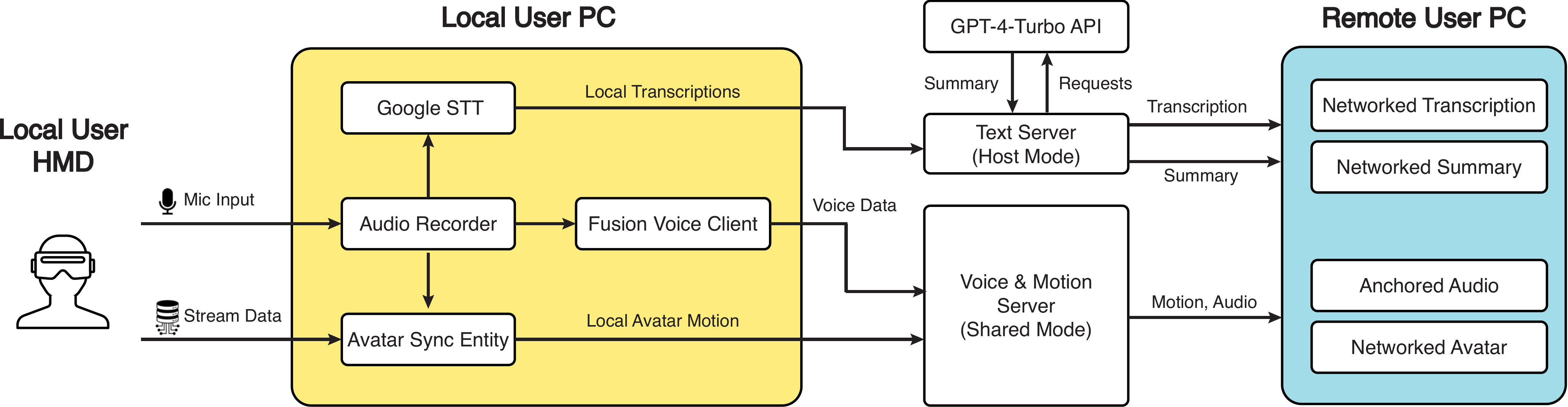}
	\vspace{-1.0em}
	\caption{
    Real-time multi-user networking pipeline for our immersive VR meeting setup. Local microphone inputs are processed by Google STT for transcription, while audio streams to a shared server via Fusion Voice Client for voice and avatar synchronization. The Text Server (host mode) manages transcriptions and requests summaries from GPT-4-Turbo API. All elements (transcription, summaries, audio) synchronize across remote users to maintain seamless real-time interaction}
	\label{fig:network_pipeline}
\end{figure*}

\section{Formative Study}
To explore how users interact with transcription interfaces in immersive VR meetings, we conducted a formative study examining behavior during disruptions and re-engagement. The study focused on two key scenarios: i) participants' reliance on live transcriptions during discussions, and ii) their ability to catch up on missed content after a period of absence. We compared two transcription interfaces: a full transcription mode, where all speech was transcribed, and a summarized version, where each utterance (defined as non-stop sentences spoken by a single user) was condensed to approximately 10 words. This builds on prior research emphasizing the need for concise transcriptions for quick comprehension in meetings~\cite{son2023okay}, with a specific focus on investigating the use cases of each transcription mode in our drop-out and rejoin scenario.


Participants in the study wore Oculus Quest Pro HMDs and engaged in a multi-user networked meeting setup, where four users were seated in separate rooms. To simulate longer disruptions, participants were asked to remove their VR headsets and rejoin the meeting after a set interval (see Section~\ref{formative_procedure} for procedural details). 
By alternating between "dropping out" and rejoining, we evaluated how effectively each transcription method supported participants' re-engagement with the conversation.



\subsection{Real-Time Immersive VR Meeting Setup} 
Here we detail our immersive VR meeting environment setup, focusing on transcription interface and multi-user networking.
Participants, represented as avatars, communicated in real-time and were assisted by live transcriptions and summaries displayed on a table-fixed transcription interface. We implemented a VR meeting room environment similar to Meta Horizon Workrooms, with participants seated around a virtual table, using Oculus Quest Pro HMDs.

\subsubsection{Immersive Meeting Room and Transcription Interface Setup}

Using Unity 3D Engine 2022.3.8f1, we created a virtual meeting room modeled after Meta Horizon Workrooms~\cite{MetaHorizonWorkrooms}. Meta's Avatar SDK~\cite{MetaAvatarSDK} enabled upper-body tracking, lip-sync animation, and eye-blink synthesis with Oculus Quest Pro's eye-tracking. Participants could customize avatars with a mirrored selection feature that could be hidden to reduce distractions.

The transcription interface, fixed to the table in front of each participant, mirrored the UI panel setup in Meta Horizon Workrooms' remote desktop feature~\cite{MetaHorizonWorkroomsRemoteDesktop} (See Figure~\ref{fig:formative_setup}). It displayed either full transcripts or summaries, with color-coded participant names for clarity. Users navigated using up-and-down buttons or an auto-scroll option, where the interface responded when users' virtual hands intersected with the buttons.

\subsubsection{Live Speech Transcription and Summarization}
Participants' speech was transcribed using Google Speech-to-Text (STT) API~\cite{GoogleSpeechToText} to achieve high transcription accuracy and a low word error rate. Since no pre-existing Unity package was available for long-duration (>20 min) live transcription, we developed a custom Unity plugin compatible with both Windows and Android platforms. The plugin captured 48 kHz audio from VR HMD microphones, split it into segments based on pauses between words, and asynchronously processed the speech data via the API. The average word error rate from the three participant groups is 15\%.

Once transcriptions were generated, we used OpenAI GPT-4-Turbo~\cite{GPT4Turbo} to summarize each utterance, a continuous speech segment from when a person starts speaking until they pause, within a 10-word limit. 
Two researchers analyzed summary accuracy against manually verified transcripts: 89\% accurately captured essential content, 9\% omitted minor details, and 2\% contained significant errors.
Latency was measured to ensure the system met the requirements for real-time interaction. Two key stages were evaluated: transcription latency (end of speech to text display) averaged 528 ms ($SD = 114 ms$), and summarization latency (text to summary display) averaged $803 ms$ ($SD = 136 ms$). The total delay from the end of speech to summary display averaged $1331 ms$. These measurements were obtained by logging timestamps for key events in the processing pipeline, including audio recording completion, transcription results, and summary generation via GPT-4-Turbo. Participants reported the latency did not hinder usability or conversation flow.

\subsubsection{Real-Time Multi-User Networking}
The networking pipeline, shown in Figure~\ref{fig:network_pipeline}, was built using Photon Fusion 2, Photon Voice, and Meta Avatar SDK to create a real-time environment for Meta avatars. We used a Photon Fusion shared mode server to synchronize avatar body movements, reducing latency across clients compared to host mode. Photon Voice handled voice streaming, while Meta Avatar SDK animated avatars using body tracking data from Oculus Quest Pro HMDs.

To handle captions and summaries, we set up a separate Photon host mode server, with the first user (moderator) acting as the host. The host server received transcriptions, generated summaries using the GPT-4-Turbo API, and shared them across clients. This separate server reduced client-side computational load, maintaining system stability over low-latency demands for caption generation. The research moderator also facilitated the session, ensuring smooth conversation flow and guiding participants during silences.

When a user’s speech was transcribed locally via Google STT, an RPC signaled the host server to generate a concise summary via GPT-4-Turbo. A data processing queue managed the order of utterance inputs, ensuring accurate token generation. The server matched each user’s Photon network ID with their respective captions and summaries, displaying them on the tabletop panel in the correct color and position.

\subsection{Participants}
We recruited $12$ participants (5 female, 7 male) for our formative study, ranging in age from $20$ to $39$ years ($M = 27.78$, $SD = 6.91$). Participants were organized into four groups, with three participants per group. 
On average, participants reported moderate VR familiarity ($M = 3.5$ on a 7-point Likert scale). All had normal or corrected vision without color blindness. Participants received \$15 e-gift cards. The university's IRB approved the study.

\subsection{Procedure} \label{formative_procedure}

Participants were grouped by availability, with three per group plus the experiment coordinator as a participant. Each participant was escorted to separate rooms on the same floor, equipped with a PC and Oculus Quest Pro, all connected to the same Wi-Fi. They calibrated their eyes using the Oculus Quest Pro’s eye calibration tool.

Participants experienced two debate-style trials per group, with both trials requiring selection of two preferred items from seven options. The first trial used the desert survival task~\cite{torre2019effect, kim2020reducing}, while the second focused on selecting optimal workplace perks. Each group experienced both full-transcript and summarized transcript interfaces, with the order counterbalanced across groups. These topics were designed to encourage differing opinions and consensus-building. The experiment coordinator facilitated the conversation and acted as an agitator if consensus was reached too quickly.

\begin{figure}[h]	
\centering\includegraphics[width=0.45\textwidth]{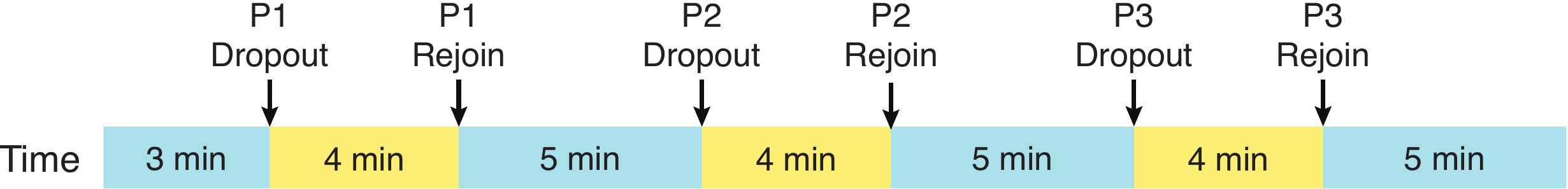}
    \vspace{-0.5em}
	\caption{An example of dropouts and rejoins of participants during the formative study. Note that the order of drop-out was randomized between trials.}
    \Description{Timeline diagram depicting the sequence of participant dropout and rejoin during the user study. The figure shows a 3-minute initial conversation period, followed by a 4-minute dropout phase where participants are interrupted by a phone chime and teleported to a different virtual environment to complete a speed math test. A timer displayed above the test indicates the time remaining until they rejoin the meeting. After the 4-minute dropout, participants rejoin for another 3-4 minutes of conversation. This diagram illustrates the study's design for simulating real-world meeting disruptions and testing the effectiveness of different transcription interfaces for re-engagement.}
	\label{fig:formative_timeline}
    \vspace{-0.5em}
\end{figure}

Before the trials began, participants were consented and given a brief tutorial on how to use the system. Once all participants were comfortable using the system, they were asked to decide on a stance for each debate topic. Once participants were ready, we proceeded to conduct the two sequential trials.
Each trial lasted about 30 minutes, during which each of the three participants dropped out of the meeting for a different four-minute interval.  Figure ~\ref{fig:formative_timeline} describes the timeline in detail, highlighting how participants took turns "\textit{dropping out}" for four minutes before rejoining. The coordinator guided the drop-out/rejoin process, and participants were asked to catch up on the conversation upon rejoining. 

\vspace{-0.5em}
\vspace{0em}

For each trial the order in which participants were chosen to drop out was selected by a random number generator.  Participants were not told in advance when they would be forced to drop out, simulating a sudden disruption such as having to answer a call or attend to a child.

During the experiment, we logged each participant's gaze behavior to assess focus and interaction with the transcription panel. 

After each trial, participants filled out questionnaires including the NASA-TLX for cognitive load~\cite{hart1988development} and  Networked Minds Social Presence Inventory (NMSPI)~\cite{harms2004internal}. We selected NMSPI factors critical for VR meetings: \textit{Co-presence} and \textit{Attentional Allocation} to measure social engagement with avatars versus interfaces, and \textit{Perceived Message/Affective Understanding} to assess conversation comprehension and emotional awareness. These NMSPI factors have been widely used in prior work evaluating social presence in virtual group discussion~\cite{kim2024engaged,wang2024socially}. Participants also rated the interfaces on effectiveness, ease of use, and preference using a 5-point Likert scale.

Semi-structured interviews were conducted after the trials to gather insights into challenges with keeping up with meetings and preferences for the transcription interfaces. The entire experiment lasted approximately 90 minutes.

\subsection{Findings}
\begin{figure*}[t]	
\centering\includegraphics[width=1.0\textwidth]{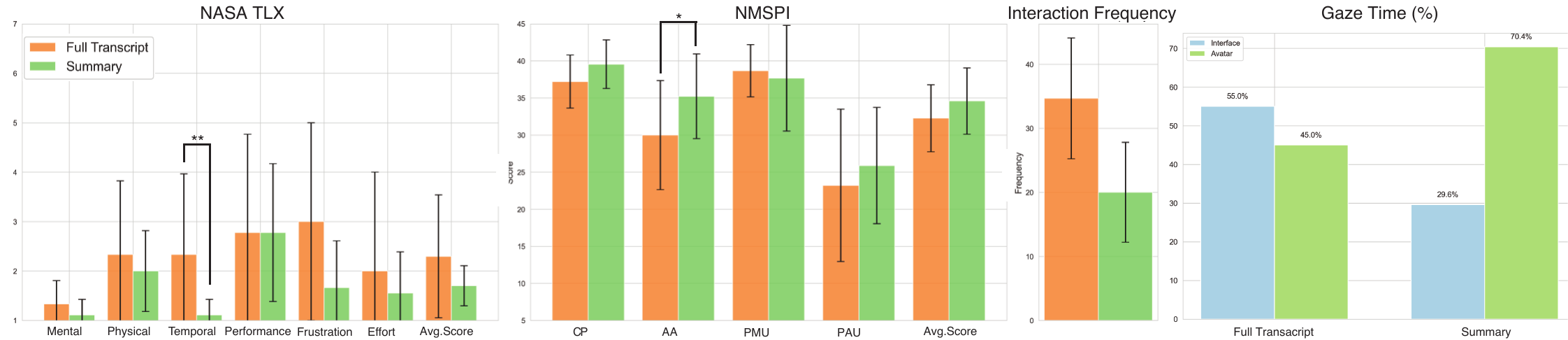}
	\vspace{-1.0em}
    	\caption{Statistical analyses from the formative study comparing \textit{Full Transcript} and \textit{Summary}. The NASA TLX results show lower cognitive load for \textit{Summary} across multiple subscales (Mental, Physical, Temporal, Frustration, and Effort) with Temporal Demand being significantly lower with \textit{Summary}. The NMSPI scores highlight significantly higher Attention Allocation (AA) for \textit{Summary}, with relatively smaller differences in Co-presence (CP), Perceived Message Understanding (PMU) and Perceived Affective Understanding (PAU). Significant differences are indicated by * (\(p < 0.05\)) and ** (\(p < 0.01\)). Participants interacted more frequently with \textit{Full Transcript}, showing similar gaze time spent between interface and avatars, whereas with \textit{Summary}, participants directed their gaze predominantly toward avatars.
    }
    \Description{Four sets of bar charts comparing NASA TLX, NMSPI scores, interaction frequency, and gaze time for Full Transcript and Summary interfaces from the formative study. The NASA TLX results show lower cognitive load for the Summary interface across multiple subscales (Mental, Physical, Temporal, Effort, Performance, and Frustration). The NMSPI scores highlight significantly higher Attention Allocation (AA) for the Summary interface, with relatively smaller differences in  Co-presence (CP), Perceived Message Understanding (PMU) ,and Perceived Affective Understanding (PAU). Participants interacted more frequently with the Full Transcript interface, showing similar gaze time spent between interface and avatars, whereas with Summary, participants directed their gaze predominantly toward avatars.}
	\label{fig:formativestat}
\end{figure*}

In this section, we report qualitative and quantitative findings from our formative study, using F1–F12 to reference participants in qualitative feedback. The interviews were audio-recorded, transcribed, and iteratively coded by two researchers to identify key themes. The statistical figure is shown in Figure~\ref{fig:formativestat}.

\paragraph{Gaze Time and Interaction Frequency}
Two researchers double-coded the gaze and interaction data, cross-checking with video recordings to confirm or adjust for hardware errors.
Analysis of gaze time and interaction frequency revealed notable differences between full-transcript and summary modes. 
Participants spent more time gazing at the text panel in full-transcript mode ($M = 142.67s, 55.6\%$ of total time), often at the expense of engaging with other participants. In contrast, summary mode reduced gaze time ($M = 72.67s, 29.6\%$ of total time) on the tabletop, allowing for more interaction with the group. F7 noted, "\textit{It can be a little distracting when there's so many words... Sometimes, I just look at the texts instead of looking at the people}". 
Similarly, the frequency of interactions with the panel was lower in summary mode ($M = 20$ times) compared to full-transcript mode ($M = 34.67$ times).

\paragraph{Cognitive Load}
Cognitive load, measured using the NASA-TLX, showed no significant differences between the modes in mental demand, physical demand, effort, frustration, or overall score. However, a paired t-test revealed a significant difference in temporal demand($p = .032$), with summary mode outperforming full-transcript mode. This suggests that participants felt less time pressure using the summary interface. 

\paragraph{Social Presence}
Although no significant differences were found in co-presence, perceived message understanding, or perceived affective understanding between the two modes, summary mode had higher average scores across all factors. The only statistically significant difference was in attentional allocation, with summary mode scoring higher (M = 40.5, SD = 6.08) than full-transcript mode (M = 30.0, SD = 6.43), indicating participants felt their attention was better distributed in summary mode ($p < .01$)(see Figure~\ref{fig:formativestat} right).

\paragraph{User Preferences}
Using a 5-point Likert scale, both modes were rated similarly in perceived usefulness for keeping up with meetings (summary: 4.44, full-transcript: 4.0). However, 9 out of 12 participants preferred the summary mode in a forced choice evaluation. Those who preferred the full-transcript mode wanted more detail and to follow the entire conversation, which some found most useful when following up with current speakers (F1, F6, F8, F11). In contrast, those favoring summary mode felt it was more helpful when they were catching up with what they had missed during the conversation and when rejoining after dropping out. Some noted that transcription errors, compounded by summarization, could occasionally result in loss of context (F8, F11).

\paragraph{Challenges}
Participants highlighted several challenges with the transcription panel. Many found that the inability to capture tones and gestures hindered their understanding of nuanced conversations (F1, F6, F10, F12). Others found it difficult to catch up when asked for input while still reviewing missed content (F1, F2, F5). Frequent gaze switching between the panel and other participants was also a common issue, with some noting it was more disruptive than in traditional platforms like Zoom (F2). Several participants mentioned that scrolling through long transcripts felt overwhelming (F4, F8), leading some to skim for keywords rather than fully reading the text (F4). Concerns were raised that these issues might be more pronounced in larger groups or more complex discussions (F2, F5, F7, F8, F9).

\paragraph{Desired Features}
Participants also provided valuable insights into how the transcription panel could be improved to better support their needs during immersive meetings. Their suggestions focused on enhancing the usability of the interface, allowing for more efficient re-engagement with the conversation, and better integrating the panel with the spatial dynamics of the VR environment.

Key areas for improvement highlighted by participants include:
\begin{itemize}
    \item The ability to catch up on missed content at a glance, enabling quicker re-engagement (F1, F2, F9).
    \item An option to attach opinions or key points directly next to the respective participant’s avatar, enhancing contextual understanding (F1, F2, F7, F8).
    \item Participants highlighted the need for greater utility in larger group meetings, where the interface could help manage more complex discussions (F1, F2, F7).
    \item There was an emphasis on improving the spatial integration of the transcription panel within the immersive environment to enhance the overall meeting experience (F2, F4, F10, F12).

    \item A feature to disable the transcription panel when not needed, as some participants found it distracting (F5, F7, F8, F10, F11).
    \item The ability to switch between full-transcript and summary modes based on the context and user preference (F2, F3, F5, F7, F11, F12).
\end{itemize}

\subsection{Design Considerations}
Drawing from our formative study findings, we propose three design considerations (DCs) to guide the development of the transcription panel for immersive meetings. Two researchers used an affinity diagramming process to organize these findings and identify recurring themes.

\paragraph{\textbf{DC1: Attach transcripts to avatars to reduce distraction.}}
Participants often found the fixed transcription interface distracting, which reduced social presence and attentional allocation. The full-transcript mode, in particular, increased temporal demand as users felt pressure to keep up with the continuous text, leading to disengagement. F11 noted, \textit{"I felt the rush to read through it quickly… I noticed others were not paying attention, just interacting with the transcription interface in front of them."} Eight participants expressed a preference for avatar-fixed transcripts, which would help them balance attention between the interface and group conversation. Similar to how prior work observed DHH users preferring captions placed near speakers in group settings~\cite{jain2018exploring}, \textbf{DC1} highlights the benefits of avatar-fixed transcripts for reducing distractions and supporting social engagement. 

\begin{figure*}[t]
	\centering
    \includegraphics[width=1.0\textwidth]{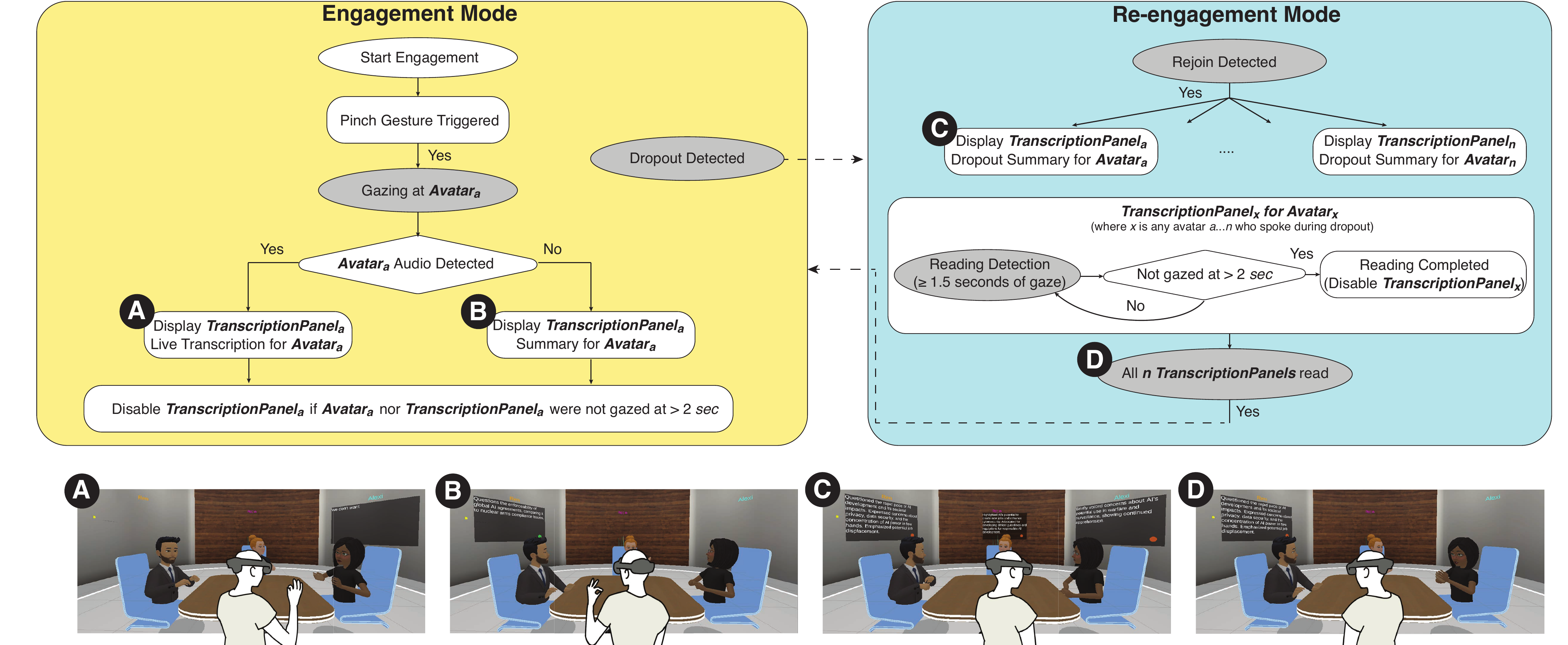}
	\vspace{-1.5em}
	\caption{An overview of \textbf{EngageSync} flowchart and demonstrational screenshots of key features. In Engagement Mode, (a) if the user performs a pinch gesture while looking at a speaker, the panel attached to the avatar displays a live transcription; (b) if the avatar is a listener (no audio detected), a summary of their previous utterance is shown. Upon rejoining after a dropout, (c) summaries of what each avatar said during the dropout are displayed. (d) Once it is `read', the interface disappears, and when all the summary panels are read, the system returns to Engagement Mode.}
    \Description{Comprehensive flowchart depicting the operation of EngageSync in both Engagement and Re-engagement modes. In Engagement Mode, the system responds to user gaze and pinch gestures, displaying live transcriptions for speakers and summaries for listeners. The Re-engagement Mode is triggered upon user dropout, providing summaries of missed conversations when the user returns. The flowchart illustrates decision points based on gaze detection, audio activity, and reading completion, showing how EngageSync dynamically adapts to user behavior and meeting context. This visual representation demonstrates the system's logic for maintaining engagement and facilitating smooth re-engagement in immersive VR meetings.}
	\label{fig:engagesync}
\end{figure*}

\paragraph{\textbf{DC2: Adapt transcription modes based on user context.}}
Preferences for transcription modes varied depending on the meeting context. During an ongoing conversation, participants preferred full transcripts to catch up, but for reviewing past discussions, they favored summaries for quicker re-engagement. They wanted the system to be aware of their disengagement periods and tailor summaries accordingly. This aligns with gaze pattern analysis, showing varied interactions based on engagement level. Thus, \textbf{DC2} suggests adaptive transcription modes that switch between full and summarized views based on user context to optimize information delivery.

\paragraph{\textbf{DC3: Provide on-demand access to the transcription interface.}}
Several participants found the constant presence of the transcription interface distracting during conversations. F3 and F12 mentioned that being able to toggle the interface on and off would reduce cognitive load and help maintain focus. On-demand access would allow users to manage distractions and avoid interference with the conversation flow. This shows similarity to findings from prior work~\cite{rivu2020stare}, which discussed users' desire to access information on demand with AR information overlays. Therefore, \textbf{DC3} focuses on designing the interface to offer on-demand access, letting users control when to engage with the transcription information and ensuring it remains a supportive rather than intrusive element. 

\vspace{1.0em}
These DCs reflect the nuanced challenges and opportunities uncovered in our study of live, multi-user VR discussions. While some findings resonate with prior work on caption placement and on-demand interfaces, our study highlights the importance of tailoring these elements to the specific demands of immersive group meeting settings, particularly in scenarios involving disruptions and re-engagement.

\section{EngageSync}

Building on the design considerations (DCs), we developed \textbf{EngageSync}, a transcription interface that enhances social presence and information recall in immersive meetings by dynamically adapting information on avatar-fixed text panels. 

\subsection{System Overview}

EngageSync adapts content based on user \textbf{\textit{context}}, referring to their engagement and attention state in the meeting. Our formative study revealed that users have different interaction strategies and preferences with transcription panels when focused on speakers, listeners, or when re-engaging after disengagement.
The system defines three key \textbf{\textit{contexts}} based on user behavior:
\begin{itemize}
    \item \textbf{Focused on a Speaker}: When the user directs their gaze at a speaking avatar, this indicates active engagement with the conversation.
    \item \textbf{Focused on a Listener}: When the user focuses on a listening avatar (i.e., an avatar that is not actively speaking), it reflects the user’s attention on non-verbal aspects of the meeting.
    \item \textbf{Re-engaging after Disengagement}: If the user’s gaze is not focused on any avatar or object in the VE, the system detects disengagement. Once the user gaze is detected on the avatars again, the system recognizes a re-engagement scenario.
\end{itemize}

 To ensure users can easily differentiate between the modes, each text panel is marked with a colored circle in the bottom-right corner. Live-transcription panels do not feature a circle, engagement summaries display a green circle, and re-engagement summaries display an orange circle. This visual indicator helps users quickly identify the current mode, reducing confusion and improving interaction flow (see Figure~\ref{fig:teaser} for visual examples).

\textbf{Gaze Tracking and Speech Detection:} EngageSync uses gaze tracking and speech activity detection to monitor user engagement and context. The Meta Movement SDK detects the user’s gaze direction, determining which avatar they are focused on. Simultaneously, speech activity detection determines whether the avatar being observed is speaking or listening. Based on this input, the system toggles between live transcription and summary modes to ensure users receive the appropriate content. For instance:
\begin{itemize}
    \item When the user focuses on a speaking avatar, EngageSync displays a live transcription panel attached to that avatar, providing real-time updates of the conversation.
    \item When the user focuses on a listening avatar, the system displays a summary of that avatar’s last utterance, offering a brief overview rather than continuous text.
\end{itemize}

\begin{figure*}[h]	
\centering\includegraphics[width=1.0\textwidth]{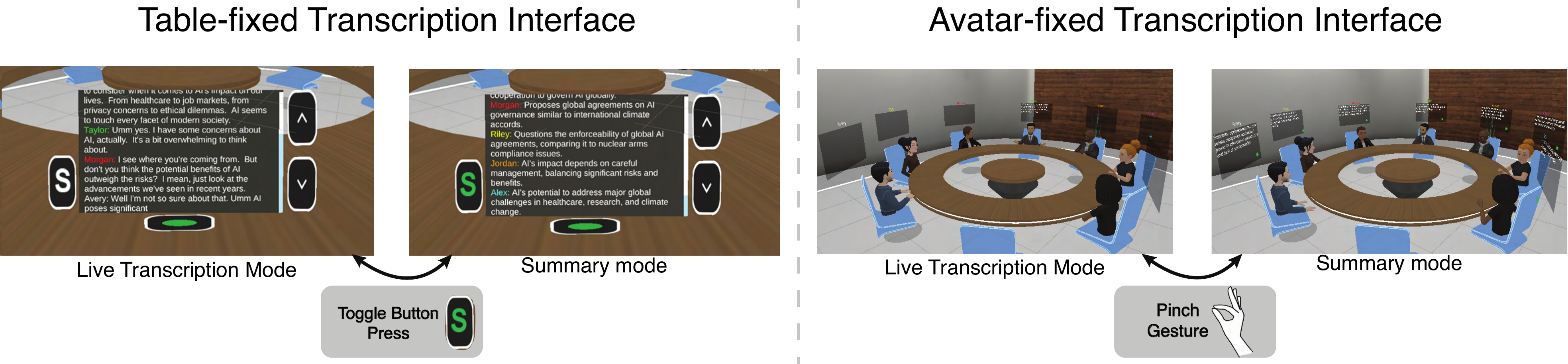}
	\vspace{-2.0em}
	\caption{The two interface compared with EngageSync in the user study. Users can toggle between live transcription mode and summary mode for both interfaces. For Table-fixed Transcription Interface (Left), this is triggered by simply pushing the 'S' button on the interface. In Avatar-fixed Transcription Interface (Right), users toggle between mode by doing a pinch gesture regardless of their gaze.}
	\vspace{-1em}\label{fig:compared_interface}
\end{figure*}

This approach to context-driven transcription directly addresses user feedback from the formative study, where participants noted that continuously receiving transcription while not focusing on the speaker could become overwhelming.

\textbf{Gesture Interaction:} To provide more intuitive control over transcription access, EngageSync leverages gesture-based interaction through the OVR Interaction SDK. Users can perform a pinch gesture while looking at an avatar to activate the corresponding transcription panel. This interaction method allows the user to summon or dismiss the transcription interface as needed, reducing cognitive load when it’s not necessary.

Once activated, the transcription panel remains visible until the system detects a lack of gaze focus (i.e., if the user looks away or is inactive for over two seconds), at which point the transcription panel automatically fades, addressing the need for on-demand access (\textbf{DC3}). This gesture-based control enables users to balance information retrieval and social presence during the meeting.

By combining gaze tracking, speech detection, and gesture interaction, EngageSync dynamically adapts the information presented to users based on the current context. This context-aware approach allows users to stay engaged in the conversation without being distracted by unnecessary or overwhelming information. The system’s flexibility in presenting full transcriptions during active engagement or concise summaries during re-engagement ensures that users can maintain their social presence in the meeting, even after disruptions.

\subsection{Interaction Workflow}

\textbf{Engagement Mode} manages user behavior during active meeting participation. Through gaze tracking and speech detection, EngageSync dynamically adapts transcription content based on user focus.
When users look at a speaking avatar, a live transcription panel appears attached to that avatar. For listening avatars, the system displays a summary of their last utterance instead. The transcription panel automatically disappears after two seconds without user focus, aligning with \textbf{DC1}. Users can also control panel visibility through pinch gestures, reducing cognitive load.

\textbf{Re-engagement Mode} activates when users temporarily disengage from the meeting, such as by looking away or becoming inactive. The system detects disengagement through the absence of gaze on avatars or objects, logging ongoing conversations during this period.
Upon re-engagement (the user's gaze is detected on avatars or virtual objects), the system generates 15-word summaries of missed conversations using GPT-4 Turbo. These summaries are attached to relevant avatars, facilitating quick catch-up. The system tracks summary engagement through gaze duration; panels disappear once read (after 1.5 seconds of focus), unless the user looks back within two seconds. After all summaries are reviewed, the system returns to Engagement Mode. For a visual overview, refer to Figure~\ref{fig:engagesync}.

When disengagement is detected, the system records ongoing conversations, capturing full utterances from speaking avatars to provide users with complete context. Upon re-engagement—, the user's gaze refocuses on avatars or virtual objects—concise 15-word summaries of the missed conversations are generated using GPT-4 Turbo. These summaries are attached to the corresponding avatars, helping the user quickly catch up on key points.

The system tracks whether users engage with these summaries by monitoring their gaze. If the user’s gaze remains on a summary panel for more than 1.5 seconds, it indicates that they have begun reading the panel. Once a summary is read, the panel disappears, unless the user looks back within two seconds, allowing the information to be reviewed without overwhelming the user with excessive content.

After all summaries have been reviewed, the system automatically transitions back to Engagement Mode, resuming real-time transcription and summaries based on the user’s focus. This adaptive interaction flow ensures that users remain informed and engaged, even after temporary disruptions, maintaining their social presence in the meeting without unnecessary distractions.
\section{User Study}
To evaluate the effectiveness of EngageSync, we conducted a user study comparing it against two other transcription interfaces in immersive VR meetings.

\begin{table*}[h]
\centering
\begin{tabular}{c|c|c|c}
\Xhline{3\arrayrulewidth}
\textbf{Feature} & \textbf{Table TI} & \textbf{Avatar TI} & \textbf{ES (EngageSync)} \\
\Xhline{3\arrayrulewidth}
\textbf{Panel Position} & Fixed on Table & Fixed to Avatar & Fixed to Avatar \\
\hline
\textbf{Missed Content Handling} & Yes (Manual Scroll) & No & Yes (Automatic Summary) \\
\hline
\textbf{Activation} & Always Visible & Always Visible & Gaze and Gesture-Triggered \\
\hline
\textbf{Transcription Mode Switching} & Manual (Button) & Manual (Gesture) & Automatic (Context-based) \\
\Xhline{3\arrayrulewidth}
\end{tabular}
\caption{Comparison of transcription panel configurations. EngageSync shares common aspects with Avatar TI in being avatar-fixed, and with Table TI in handling missed content. However, EngageSync stands out by offering on-demand access (gaze and gesture-triggered) and automatic transcription mode switching, making it more adaptable and less intrusive in immersive meetings.}
\label{table:comparison}
\vspace{-1.5em}
\end{table*}

\subsection{Compared Interfaces and Hypotheses}
In our study, we compare EngageSync to two other transcription panel configurations: Table-fixed Transcription Interface and Avatar-fixed Transcription Interface. See Figure~\ref{fig:compared_interface} for visual reference.

\textbf{Table-fixed Transcription Interface (TableTI)}: This interface, expanding from the version used in our formative study, positions the transcription panel in front of the user in a fixed position on the table. We introduced a toggle button to the left of the text panel to switch between real-time transcription and a summary of each speaker's previous utterances. The interface retains the familiar scrolling controls with up and down buttons and an auto-scroll option to the right and bottom. When a user rejoins the meeting after a disruption, the text display resumes from where they left off, ensuring continuity.

\textbf{Avatar-fixed Transcription Interface (AvatarTI)}: This interface attaches transcription panels directly above each avatar, similar to live captions or subtitles used in social VR platforms like VRChat~\cite{VRChat}. These panels are constantly visible, and the user can toggle between live transcription mode and summary mode using a pinch gesture. In live transcription mode, only the currently speaking avatar has its text displayed, while in summary mode, all avatars display a summary of their previous utterance. This interface does not adapt to disengagement and simply shows previous summaries, as common avatar-attached interfaces typically lack context awareness. 

On the other hand, EngageSync adapts the display of transcription panels based on the user’s engagement. Instead of always showing the panels, users must look at an avatar and make a pinch gesture to activate the panel. The content displayed depends on the user's context, providing live transcriptions when engaged and summarizing missed conversations when re-engaging after a disruption.
The differences and common aspects between these interfaces are summarized in Table~\ref{table:comparison}.

Our study aims to test the following hypotheses, based on findings from the formative study and previous research:

\begin{itemize}
    \item \textbf{H1: EngageSync provides the best overall balance between social presence and the ability to keep up with the conversation.} 
    We hypothesize that EngageSync will enhance social presence by attaching transcripts to avatars (\textbf{DC1}), which directs users’ attention toward others and reduces distractions compared to TableTI. Additionally, EngageSync’s context-based summaries and live transcription (\textbf{DC2}) are expected to better meet users’ informational needs, supporting more effective re-engagement than AvatarTI.

    \item \textbf{H2: Avatar-fixed transcription interfaces increase social presence compared to baseline conditions.}
    We expect that both AvatarTI and EngageSync will provide a greater sense of social presence than TableTI. By increasing users’ gaze toward avatars, these interfaces enable better perception of others as well as engagement, which participants in the formative study noted were severely hindered with the table-fixed transcription panel.

    \item \textbf{H3: Table TI and EngageSync support better re-engagement after disengagements.} 
    We hypothesize that both TableTI and EngageSync will enable participants to catch up more effectively after disengagements compared to AvatarTI. By providing a record of prior conversation, these interfaces are expected to enhance information recall and facilitate re-engagement, addressing a key challenge identified in the formative study, where participants noted the difficulty of reconnecting with the conversation flow after disruptions.

    \item \textbf{H4: Avatar-fixed transcription interfaces are more effective in larger groups.} 
    We hypothesize that AvatarTI and EngageSync will demonstrate greater usability and effectiveness in larger groups compared to TableTI. This hypothesis is based on feedback from the formative study, where participants highlighted the need for greater utility in larger group meetings, stating that the interface could help manage more complex group dynamics.

\end{itemize}

\subsection{Experiment Design}

To test these hypotheses, we designed the user study to consist of two groups of immersive meeting setups of different sizes. Here, participants joined pre-recorded group conversations as listeners to ensure a consistent experience across all conditions and isolate the effects of our interface designs without introducing variability from differing conversation quality and disruption timing. This aligns with prior works that tested meeting interfaces in VR~\cite{lee2024may} and transcription interfaces in video meetings~\cite{son2023okay} using pre-recorded setups.

\paragraph{\textbf{Group Size Conditions}} The evaluation of these interfaces will be conducted under two distinct group sizes.

\begin{itemize}
    \item \textbf{Small Group Conversation (3 speakers):} Following the setup of our formative study, this condition will consist of \textit{\textbf{three}} virtual agents conversing. This will result in four avatars placed in the room, including the participants.
    \item \textbf{Mid-sized Group Conversation (7 speakers):} In this condition, the group size is scaled up with participants observing a conversation among \textit{\textbf{seven}} virtual agents. This design choice was driven by formative study feedback, where users expressed that group size would affect the usability of the interface and that they would find avatar-fixed panels even more useful in larger groups.
\end{itemize}

To clarify, the study involves comparing the three interfaces within each group size condition, resulting in a within-subjects design for each group size. Additionally, the comparison between the two group sizes will follow a between-subjects design, assessing how group size influences the usability and effectiveness of the interfaces.

\begin{figure}[t]	
\centering\includegraphics[width=0.5\textwidth]{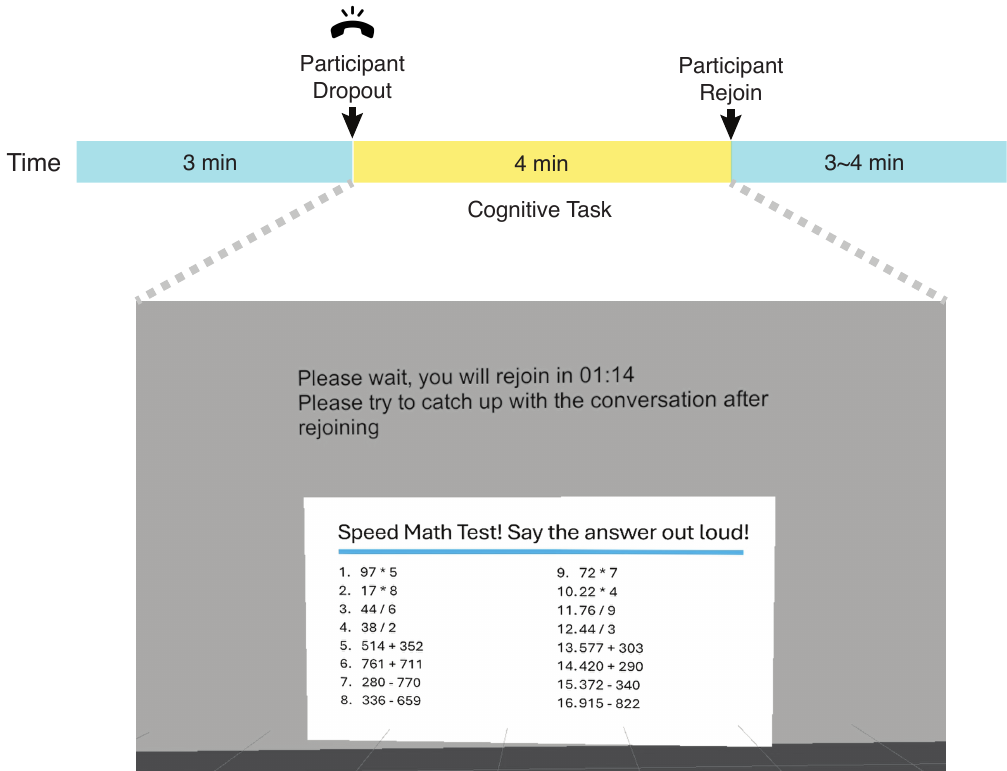}
	\vspace{-1.0em}
	\caption{
   Participant dropout and rejoin timeline. Three minutes into the conversation, a phone chime interrupts participants, who are teleported to a separate VE for a speed math test with a timer displaying the time remaining until they rejoin the meeting.
    }
	\label{fig:eval_timeline}
\end{figure}

\paragraph{\textbf{Dropout Simulation}} 
To simulate real-world disruptions, the participant hears a phone ringing sound 3 minutes into the experiment, signaling a ``drop out" situation. Immediately after, the participant’s avatar is relocated to a separate virtual environment away from the meeting.  To further simulate the mental load of a disruption a simple math quiz is displayed on a large text panel (see Figure~\ref{fig:eval_timeline}). A countdown timer is also shown, indicating the time remaining until the participant is automatically returned to the meeting. Participants are instructed to solve as many math problems as possible within the remaining time. This design follows prior work~\cite{srivastava2021mitigating} that employs cognitive tasks that interrupt participants' engagement from the original task; in this case, distracting them from the previous conversation context. After 4 minutes, the participant’s avatar is relocated back to the ongoing meeting, with the conversation continuing for another 3 to 4 minutes, as each conversation script is designed to last between 10 and 11 minutes.

\paragraph{\textbf{Conversation Script}} \label{par:script}It is important to note that the script for both group size conditions is exactly the same. The speech content is divided among three avatars in the small group and seven avatars in the mid-sized group. We designed the script carefully so that a script for one speaker in the small group was distributed to three speakers in the mid-sized group. Additionally, one speaker in both groups speaks less than the others, this design was made to evaluate whether users can recall or remember what that speaker said using the different interfaces. An example of the script distribution can be found in Table~\ref{table:script_role}.

We designed three different topics/scripts for the conversations: ``\textit{Is Online Education as Effective as Traditional In-Person Education?}", 
``\textit{Should Social Media Platforms be Regulated by the Government?}", 
``\textit{Is Artificial Intelligence More Beneficial or Harmful to Society?}". 

\begin{table}[t]
\centering
\small
\setlength{\tabcolsep}{6pt} 


\begin{tabular}{>{\columncolor[HTML]{FFFFFF}}c|c|c} 
\Xhline{3\arrayrulewidth}
{\textbf{Role}} & {\textbf{Small Group (SA1-3)}} & {\textbf{Mid-sized Group (MA1-7)}} \\
\Xhline{3\arrayrulewidth}
Pro-topic & SA1 & MA1, MA4, MA7 \\
\hline
Against-topic & SA2 & MA2, MA5, MA6 \\
\hline
Less talkative & SA3 & MA3 \\
\Xhline{3\arrayrulewidth}

\end{tabular}

\caption{An example of script distribution in small and mid-sized groups. Each participant in the small group corresponds to multiple participants in the mid-sized group with similar viewpoints, while one participant in each group represents a less talkative role.}
\vspace{-3.5em}
\label{table:script_role}
\end{table}

\begin{figure*}[t]
	\centering
    \includegraphics[width=\textwidth]{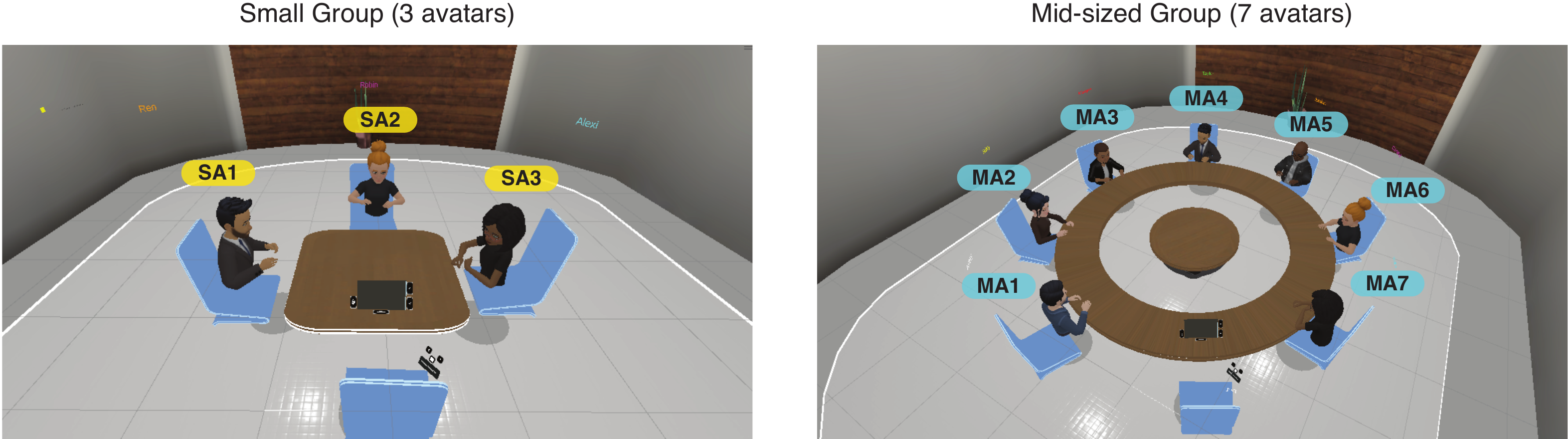}
	\vspace{-1.5em}
	\caption{The VE setup for small group with three speakers (Left) and mid-sized group with seven speakers (Right).}
	\label{fig:eval_setup}
\end{figure*}

\paragraph{\textbf{Measured Items}} 
In the study, we collected both quantitative and qualitative data to assess the effectiveness of the transcription methods in supporting user engagement and understanding in immersive meetings.

For quantitative measures, we measured the following metrics:

\begin{itemize}
    \item \textbf{Re-engagement Time:} After a trial, participants were shown a screen recording of their trial and were asked to mark the time they thought they were `caught up' with the conversation after rejoining.
    
    \item \textbf{Information Recall:} We used a two-part approach to measure information recall: a six-question quiz, and an open-ended text entry prompt asking participants to list everything they remembered from the less talkative avatar. We then coded these open-ended responses in binary form (1 if the main point was accurately captured, 0 otherwise) and  examined how well participants remembered the least talkative speaker in each trial.
    
    \item \textbf{Gaze Time:} We measured the time a participant spent gazing at the interface versus avatars in percentage out of the trial duration excluding dropout time. The gaze tracking data was automatically logged and was cross-checked with two researchers compared with screen recordings of the trial.
    
\end{itemize}

For qualitative measures, we collected the following data:

\begin{itemize}
    \item \textbf{Social Presence:} We used the NMSPI questionnaire from the formative study to evaluate social presence, focusing on co-presence, attention allocation, PMU, and PAU.
    \item \textbf{Cognitive Load:} The NASA-TLX questionnaire (7-point Likert scale) was used to measure participants' perceived cognitive load.
    \item \textbf{Usability:} Measured using the System Usability Scale (SUS)~\cite{Brooke1996SUS} questionnaire in a 5-point Likert Scale.
    \item
    \textbf{Utility Questions} We further asked participants how much the perceived interface supported them in (i) keeping up with the conversation (ii) and after re-joining in the conversation on a 5-point Likert scale. Finally, we asked participants to rate their how they prefer having this interface in an immersive meeting setup in a 5-point Likert Scale.
\end{itemize}

\subsection{Setup}
The setup for this user study largely mirrors that of our formative study, leveraging the same foundational technologies and equipment. We used Unity running on a PC equipped with 32GB RAM, an Intel i9 processor, and an NVIDIA GeForce RTX 3080 Ti. Participants wore a Meta Quest Pro headset, which supports both gesture and eye tracking. The Meta Avatar SDK was again utilized to represent avatars, maintaining consistency with the formative study.

The avatars in this user study were pre-recorded with voice actors reciting scripts specifically designed for the two group sizes: three actors for the small group and seven actors for the mid-sized group. The motion capture was performed using the Avatar Recording tool~\cite{MetaAvatarRecording}, and all conversation sequences were synchronized using Unity Timeline to ensure consistency across sessions.

The VE remained a conference room setup, with avatars seated around a table. The configuration of the environment for each group size is illustrated in Figure~\ref{fig:eval_setup}. Each speaker's name was virtually generated, color-coded, and attached above their avatar to aid identification. Additionally, for AvatarTI and EngageSync conditions, the text panels were placed above each corresponding avatar, as was done in the formative study.

\subsection{Participants}
We recruited 30 participants (14 female, 16 male) from a university sample, aged between 19 and 35 years (\textit{M} = 26.70, \textit{SD} = 4.19). Participants were randomly assigned to one of two group size conditions: the small group study (3 avatars) or the mid-sized group study (7 avatars), with 15 participants in each condition. None of the participants reported color blindness, and all had normal or corrected-to-normal vision.

On average, participants reported a moderate familiarity with virtual reality (VR), with a mean score of \textit{M} = 4.77 (\textit{SD} = 1.45) on a 7-point Likert scale (1 = not familiar at all, 7 = extremely familiar). They also indicated their frequency of attending online meetings (\textit{M} = 3.87, \textit{SD} = 0.92) on a scale from 1 (never) to 5 (daily). Additionally, participants reported their experience with either personally dropping out of meetings (\textit{M} = 2.75, \textit{SD} = 1.06) or observing others drop out (\textit{M} = 3.10, \textit{SD} = 1.05) on a scale from 1 (never) to 5 (always). This study procedure was approved by the university’s IRB for human-subject studies.
Participants were compensated with a \$15 e-gift card for their participation.



\begin{table*}[htbp]
\centering
\small
\setlength{\tabcolsep}{4pt}

\definecolor{lightpink}{rgb}{1.0, 0.9, 0.9} 
\definecolor{darkpink}{rgb}{1.0, 0.75, 0.75} 
\definecolor{deeppink}{rgb}{1.0, 0.6, 0.6}

\begin{tabular}{
    >{\centering\arraybackslash}p{2.0cm}| 
    >{\centering\arraybackslash}p{1.6cm}| 
    *{3}{>{\centering\arraybackslash}p{3.6cm}} 
}
\Xhline{3\arrayrulewidth}
\multicolumn{2}{c|}{{\textbf{Measurement}}} & \multicolumn{1}{c|}{{\textbf{Small Group}}} & \multicolumn{1}{c|}{{\textbf{Mid Group}}} & \multicolumn{1}{c|}{{\textbf{Combined Group}}} \\
\Xhline{3\arrayrulewidth}

\multirow{8}{*}{\centering Social Presence\hspace{0pt}} 
    & \multirow{2}{*}{CP} & \cellcolor{lightpink}\underline{ES > Avatar > Table} & \cellcolor{darkpink}\underline{ES > Avatar > Table} & \cellcolor{deeppink}\underline{ES > Avatar > Table} \\
    & & \cellcolor{lightpink}$\chi^2 = 6.739$  $p = .032$ $W = .225$& \cellcolor{darkpink}$\chi^2=15.500$ $p =.004$ $W = .517$& \cellcolor{deeppink}$\chi^2=14.126$ $p>.001$ $W = .471$\\
    \cline{2-5} 
    & \multirow{2}{*}{AA} & \cellcolor{lightpink}\underline{ES > Avatar > Table} & \cellcolor{darkpink}\underline{Avatar > ES > Table} & \cellcolor{deeppink}\underline{Avatar > ES > Table} \\
    & & \cellcolor{lightpink}$\chi^2=6.037$  $p=.049$ $W = .201$& \cellcolor{darkpink}$\chi^2=12.171$ $p = .002$ $W = .406$& \cellcolor{deeppink}$\chi^2=19.982$ $p<.001$ $W = .666$\\
    \cline{2-5}
    & \multirow{2}{*}{PMU} & \underline{ES > Avatar > Table} & \underline{Avatar > ES > Table} & \underline{ES > Avatar > Table} \\
    & & $\chi^2=.151$ $p=0.927$ $W = .005$& $\chi^2=1.782$ $p=0.410$ $W = .059$& $\chi^2=1.921$ $p=.252$ $W = .064$\\
    \cline{2-5}
    & \multirow{2}{*}{PAU} & \underline{Avatar > Table > ES} & \underline{ES > Avatar > Table} & \underline{Avatar > Table> ES} \\
    & & $\chi^2=.625$ $p=.269$ $W = .021$& $\chi^2=1.719$ $p=.423$ $W = .057$& $\chi^2=1.734$ $p=.420$ $W = .058$\\
\hline
\multicolumn{2}{c|}{\multirow{2}{*}{Gaze Time at Avatar (\%)}} & 
    \cellcolor{darkpink}\underline{ES > Avatar > Table} & \cellcolor{darkpink}\underline{ES > Avatar > Table} & \cellcolor{deeppink}\underline{ES > Avatar > Table} \\
\multicolumn{2}{c|}{} & 
    \cellcolor{darkpink}$\chi^2=11.831$ $p=.003$ $W = .394$& \cellcolor{darkpink}$\chi^2=9.552$ $p =.008$ $W = .318$& \cellcolor{deeppink}$\chi^2=21.043$ $p<.001$ $W = .701$\\
\hline
\multirow{6}{*}[1em]{\centering Information Recall} 
    & \multirow{2}{*}{Quiz} & \underline{ES > Table> Avatar} & \cellcolor{lightpink}\underline{ES > Table> Avatar} & \underline{ES > Table> Avatar} \\
    & & $\chi^2=2.259$ $p=.323$ $W = .075$& \cellcolor{lightpink}$\chi^2=6.882$ $p =.028$ $W = .229$& $\chi^2=3.406$ $p=.182$ $W = .114$\\
    \cline{2-5}
    & \multirow{2}{*}{\makecell{Recall Less \\ Talkative}} & \underline{ES > Table> Avatar} & \cellcolor{lightpink}\underline{ES > Table> Avatar} & \underline{ES > Table> Avatar} \\
    & & $\chi^2=Nan$ $p=Nan$ $W = Nan$& \cellcolor{lightpink}$\chi^2=6.048$ $p=.046$ $W = .202$& $\chi^2=4.478$ $p=0.107$ $W = .149$\\
    \cline{2-5}
\hline
    \multicolumn{2}{c|}{\multirow{2}{*}{Re-engagement Time}} & \cellcolor{lightpink}\underline{ES > Table> Avatar} & \cellcolor{deeppink}\underline{ES > Table > Avatar} & \cellcolor{deeppink}\underline{ES > Table> Avatar} \\
    \multicolumn{2}{c|}{} &  \cellcolor{lightpink}$\chi^2=6.773$ $p=0.047$ $W = .226$& \cellcolor{deeppink}$\chi^2=19.733$ $p < .001$ $W = .658$& \cellcolor{deeppink}$\chi^2=22.200$ $p<.001$ $W = .740$\\
\hline


\multicolumn{2}{c|}{\multirow{2}{*}{NASA TLX}} & 
    \underline{Table> ES > Avatar} & \cellcolor{darkpink}\underline{Table> ES > Avatar} & \cellcolor{darkpink}\underline{Table> ES > Avatar} \\
    \multicolumn{2}{c|}{} &  $\chi^2=3.333$ $p=.189$ $W = .111$& \cellcolor{darkpink}$\chi^2=13.119$ $p=.002$ $W = .437$& \cellcolor{darkpink}$\chi^2=14.721$ $p=.002$ $W = .491$ \\
\hline
\multicolumn{2}{c|}{\multirow{2}{*}{SUS}} & 
    \underline{ES = Table> Avatar} & \underline{ES > Avatar > Table} & \underline{ES > Avatar > Table} \\
    \multicolumn{2}{c|}{} &  $\chi^2=1.750$ $p=.196$ $W = .058$& $\chi^2=3.263$ $p=.159$ $W = .109$& $\chi^2=4.266$ $p=0.119$ $W = .142$\\
\hline
\multicolumn{2}{c|}{\multirow{2}{*}{Facilitated Catching Up}} & 
    \underline{ES > Avatar = Table} & \underline{ES > Avatar > Table} & \underline{ES > Avatar > Table} \\
    \multicolumn{2}{c|}{} &  $\chi^2=.0571$ $p=.972$ $W = .002$& $\chi^2=1.792$ $p=0.342$ $W = .060$& $\chi^2=1.942$ $p=.679$ $W = .065$\\
\hline
\multicolumn{2}{c|}{\multirow{2}{*}{Facilitated Re-engagement}} & 
    \cellcolor{darkpink}\underline{ES > Table> Avatar} & \cellcolor{deeppink}\underline{ES > Table> Avatar} & \cellcolor{deeppink}\underline{ES > Avatar > Table} \\
    \multicolumn{2}{c|}{} &  \cellcolor{darkpink}$\chi^2=8.509$ $p=.014$ $W = .284$& \cellcolor{deeppink}$\chi^2=19.304$ $p<.001$ $W = .644$ & \cellcolor{deeppink}$\chi^2=26.00$ $p<.001$ $W = .867$\\
\hline
\multicolumn{2}{c|}{\multirow{2}{*}{Preference}} & 
    \underline{ES > Table> Avatar} & \underline{ES > Avatar > Table} & \underline{ES > Avatar > Table} \\
    \multicolumn{2}{c|}{} &   $\chi^2=2.167$ $p=.339$ $W = .072$& $\chi^2=4.00$ $p=.135$ $W = .133$& $\chi^2=4.356$ $p=.113$ $W = .145$\\
\hline
\multicolumn{2}{c|}{Ranked Preference} & 
    {ES > Table > Avatar} & {ES > Avatar > Table} & {ES > Table > Avatar} \\
    
\Xhline{3\arrayrulewidth}
\end{tabular}
\caption{Friedman test results for different groups and measures (EngagementSync $=$ ES, TableTI $=$ Table, AvatarTI $=$ Avatar). Results with significant differences are highlighted with a \textcolor{lightpinkt}{light pink color} for $p<.05$, a \textcolor{darkpinkt}{pink color} for $p<.01$, and a \textcolor{deeppinkt}{darker pink color} for $p<.001$. Overall, EngageSync provides higher co-presence and attention allocation with faster re-engagement time with a more significant difference with mid-sized group. Kendall's W values indicate effect size magnitude where $W \approx  0.1$ is small, $W \approx  0.3$ is moderate, and $W \approx  0.5$ is large effect.}
\label{tab:statistical_results}
\vspace{-1.5em}
\end{table*}
 
\subsection{Procedure}
Participants first reviewed and signed a consent form, after which they completed a demographic survey. They were briefed on the study's procedure of experiencing a simulated drop-out. Eye-tracking calibration was performed using Meta Quest Pro’s internal eye-tracking calibration software to ensure accuracy throughout the experiment.

Each participant completed three trials, where the topic order remained fixed but the type of transcription interface provided was randomized in a counterbalanced order across participants  Before each trial, participants received a 5-minute training session on the interface they would be using for that particular trial. Once the participant felt comfortable with the interface, the trial commenced.

Each trial consisted of a 10-minute conversation, including a 4-minute "drop-out" phase. During the drop-out phase, participants were relocated to a different VE, where they were tasked with solving a simple math quiz. Participants were instructed to focus on solving the quiz while keeping an eye on the timer to be aware of the time remaining before rejoining the conversation.

After each trial, participants completed the  NMSPI, NASA TLX, SUS, and a custom set of utility questions designed to assess their experience with the interface. Upon completing all three trials, participants were asked to rank their preference among the three transcription interfaces. A semi-structured interview followed, aimed at gaining deeper insights into their experiences and preferences.

Participants were offered short breaks between trials if needed, and the entire study lasted approximately 60 to 90 minutes.

\subsection{Results}
We present qualitative and quantitative results for small and mid-sized groups, analyzing differences between these group sizes. Additionally, we evaluate the overall effect of each measure by combining all group conditions, referred to as the ``Combined Group."
The Shapiro–Wilk test indicated that the data were not normally distributed ($p < .05$), so we applied non-parametric tests throughout. Friedman tests were used to assess differences within each group condition, followed by Dunn’s post-hoc tests with Bonferroni correction for pairwise comparisons. Effect sizes are reported using Kendall’s W to quantify the degree of agreement among conditions. Additionally, Mann-Whitney U tests were used for between-group comparisons.
All statistical results, including $\chi^2$, $p$-values, and $W$, can be found in Table~\ref{tab:statistical_results}. In the following sections, we focus on reporting the significant findings with an emphasis on the key insights. 
For participant quotes from interviews, we refer to those from the small group as P1-P15 and those from the mid-sized group as P16-P30.

\vspace{-1em}
\subsubsection{Social Presence and Gaze Time on Avatar}  
\begin{figure*}[h]	
\centering\includegraphics[width=1.0\textwidth]{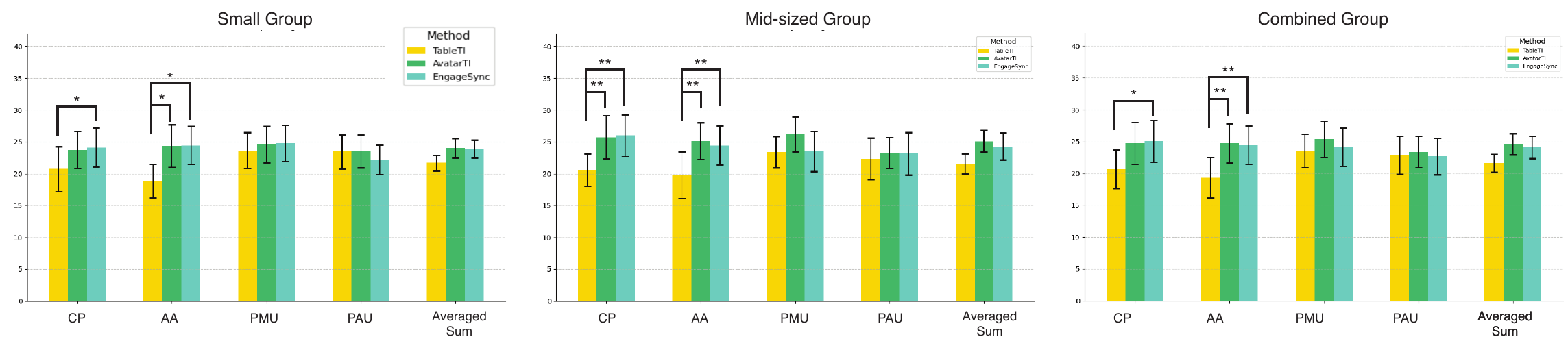}
	\vspace{-1.0em}
	\caption{
        Results for social presence across three group conditions (small, mid-sized, combined). Social presence was measured using four subfactors: Co-presence (CP), Attentional Allocation (AA), Perceived Message Understanding (PMU), and Perceived Affective Understanding (PAU), along with an averaged sum of all factors. Significant differences are indicated by * (\(p < .05\)) and ** (\(p < .01\)). Across all conditions, EngageSync consistently resulted in higher co-presence and attentional allocation scores compared to both AvatarTI and TableTI, particularly in the mid-sized group. In the combined group analysis, significant differences were observed between EngageSync and the other interfaces for CP and AA, demonstrating its effectiveness in enhancing social presence and attention management.}
	\label{fig:sp}
\end{figure*}

\hfill \break
\textbf{Social Presence:} Both EngageSync and AvatarTI, avatar-fixed transcription interfaces, yielded significantly higher CP and AA compared to TableTI across all group sizes (see figure~\ref{fig:sp}). This supports our research hypothesis \textbf{H2} that avatar-fixed transcription helps users maintain a greater sense of social presence. The effect was more pronounced in mid-sized groups, where EngageSync and AvatarTI had significantly higher CP and AA than TableTI, with large effect sizes for CP ($W = .517$) and moderate-to-large effects for AA ($W = .406$). In small groups, EngageSync and AvatarTI still outperformed TableTI in both CP and AA. 

While no significant differences were found in PMU and PAU ($p > .05$ for both group conditions), this might be due to the non-participatory nature of the study, as participants were observers, and the use of cartoonish avatars, which may have limited the expression of nuanced facial cues. Despite this, the overall improvement in CP and AA confirms that avatar-fixed systems encourage users to remain more engaged with the group rather than focusing on a table-fixed interface.

\begin{figure*}[h]	
\centering\includegraphics[width=1.0\textwidth]{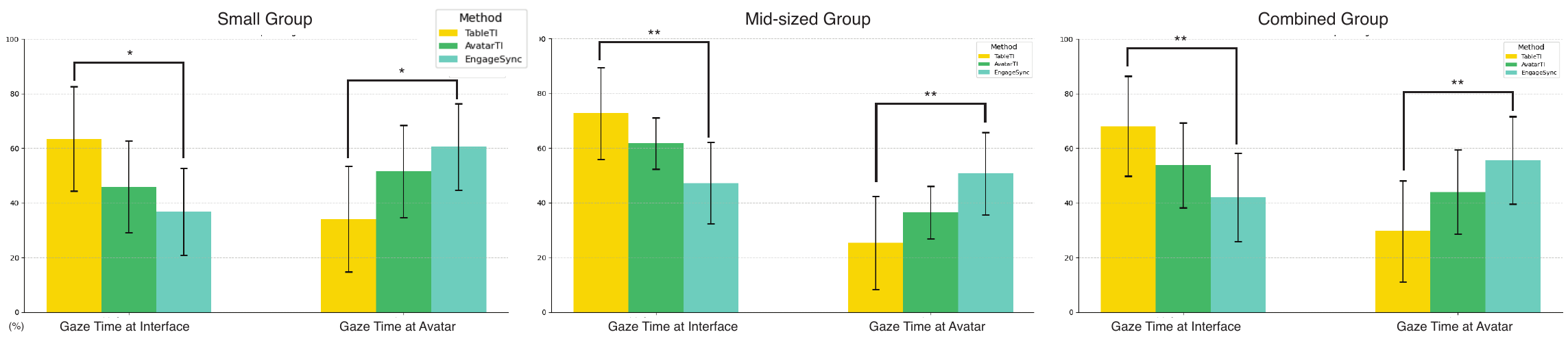}
	\vspace{-1.0em}
	\caption{
        Gaze time distribution(\%) for avatars and interfaces across different group conditions. In the small group condition, participants spent significantly more time gazing at the interface when using TableTI compared to both AvatarTI and EngageSync, whereas gaze time on avatars was higher for both avatar-fixed interfaces. A similar trend is observed in the mid-sized and combined groups, with EngageSync and AvatarTI leading to significantly more gaze time at avatars and less at the interface compared to TableTI. Significant differences are indicated by * (\(p < .05\)) and ** (\(p < .01\)). This suggests that avatar-fixed interfaces reduce user focus on the interface, enabling more natural engagement with other participants.}
	\label{fig:gaze}
\end{figure*}

\textbf{Gaze Time:} Participants spent significantly more time looking at avatars with EngageSync compared to TableTI, in both small groups ($p = .032$) and mid-sized groups ($p =.003$). No significant difference was found in gaze time between EngageSync and AvatarTI in either group size condition.

No significant differences were found between small and mid-sized groups for CP (\(U = 110.5, p = .21\)) or AA (\(U = 115.0, p = .65\)), suggesting that the benefits of avatar-fixed transcriptions remain consistent regardless of group size. However, participants in mid-sized groups spent significantly more time gazing at the table-fixed interface (\(U = 177.0, p =.0079\)) compared to small group participants, suggesting that as the conversation becomes more complex, static transcription panels become more distracting.

These findings suggest that EngageSync and AvatarTI improve social presence by allowing users to focus more on avatars rather than transcription interfaces. The results also suggest that TableTI may lead to increased distractions in larger group meetings, supporting the need for adaptive, context-aware systems. While EngageSync shows promising potential, further research is needed to fully understand how minimizing interface-gazing affects social engagement over time.

\subsubsection{Information Recall and Re-engagement Time}

\begin{figure*}[h]	
\centering\includegraphics[width=1.0\textwidth]{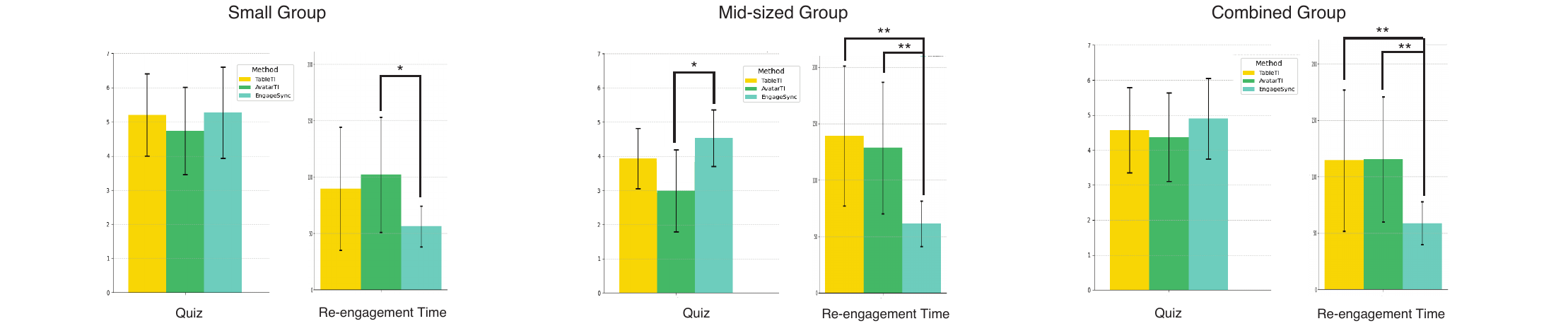}
	\vspace{-2.0em}
	\caption{Information recall results (quiz scores), and re-engagement time across small, mid-sized, and combined groups. In the small group, EngageSync significantly reduced re-engagement time compared to AvatarTI, with no significant differences in quiz scores. In the mid-sized group, EngageSync outperformed AvatarTI and TableTI in re-engagement time, and AvatarTI in quiz scores. Across the combined group, EngageSync consistently led to faster re-engagement time. Significant differences are marked by * (\(p < .05\)) and ** (\(p < .01\)).}
	\label{fig:recall}
\end{figure*}

Information recall and re-engagement time both measure how effectively participants recover from conversation interruptions. We used a six-question quiz and an open-ended text entry task (recalling less talkative speaker) to assess information recall, while re-engagement time captured how quickly participants resumed the conversation flow after being disengaged. Together, these measures indicate how well each interface supports users in regaining context, thereby informing \textbf{H3} (see Figure~\ref{fig:recall}).

\begin{figure*}[h]	
\centering\includegraphics[width=1.0\textwidth]{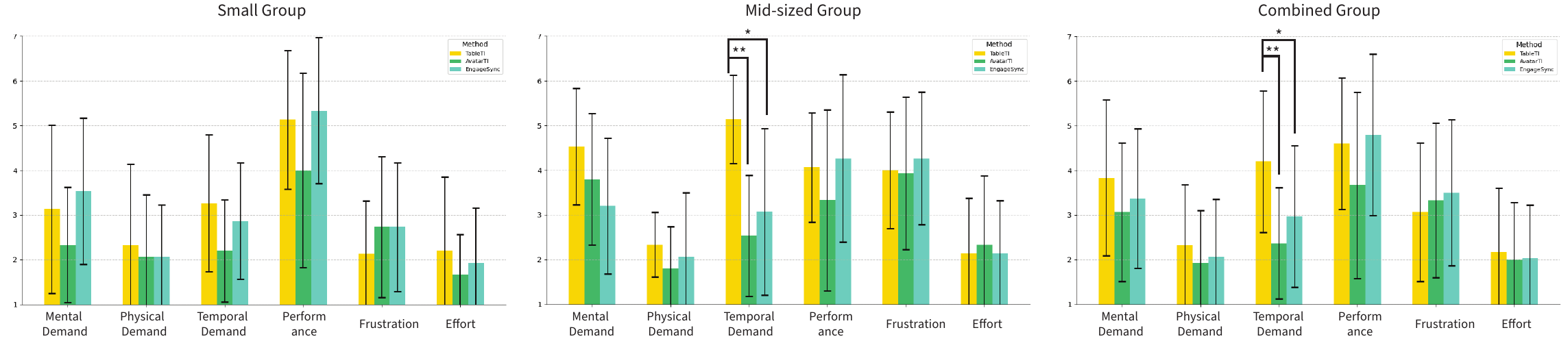}
	\vspace{-1.0em}
	\caption{
    NASA TLX results across group conditions. Temporal demand was significantly higher for TableTI compared to both AvatarTI and EngageSync in the mid-sized group, while no significant differences were observed for other subscales in the small group. The combined group analysis similarly showed higher temporal demand for TableTI, with EngageSync and AvatarTI resulting in lower mental and physical demand, frustration, and effort across all group sizes. Significant differences are indicated by * (\(p < .05\)) and ** (\(p < .01\)).}
	\label{fig:nasa}
\end{figure*}

\textbf{Information Recall:}
EngageSync consistently outperformed both TableTI and AvatarTI in supporting information recall, particularly in mid-sized groups. Although no significant differences emerged in the small group condition ($p > .05$), participants achieved significantly higher quiz scores with EngageSync than with AvatarTI in mid-sized groups. This advantage likely stems from the dynamic summaries generated during re-engagement, which helped them catch up on missed key points.

Regarding recall of the least talkative speaker’s comments, there were no significant differences in the small group condition, as all participants remembered that speaker’s opinion across all methods. In the mid-sized group ($p < .05$), however, participants in six TableTI trials and three AvatarTI trials failed to recall the least talkative speaker’s comments, whereas no such lapses occurred with EngageSync. This finding implies how EngageSync’s adaptive summaries preserve quieter voices in more crowded group discussions.

\textbf{Re-engagement Time.}
Across both group sizes, participants re-engaged with the conversation more quickly when using EngageSync than when using TableTI or AvatarTI, with this difference was more pronounced in mid-sized groups (W = .658). The flexibility of EngageSync's automatic summary display upon rejoining helped participants quickly digest what they had missed. In contrast, users of AvatarTI often reported taking longer to catch up, particularly in larger groups, where more speakers made it harder to stay oriented. These findings underscore the importance of adaptive summaries in more complex discussions, where the ability to rapidly recapture key points becomes increasingly critical for efficient re-engagement.

\begin{figure*}[h]	
\centering\includegraphics[width=0.6\textwidth]{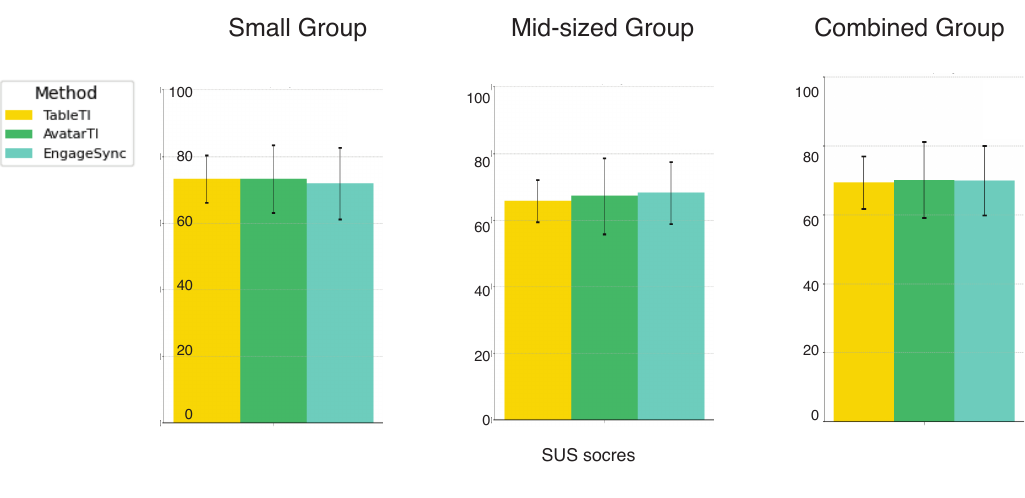}
	\vspace{-1.0em}
	\caption{SUS results across group conditions. The scores did not show any significant differences in any group, indicating that participants found all interfaces to have comparable usability, regardless of group size or interface type.}
	\label{fig:sus}
\end{figure*}
\begin{figure*}[h]	
\centering\includegraphics[width=1.0\textwidth]{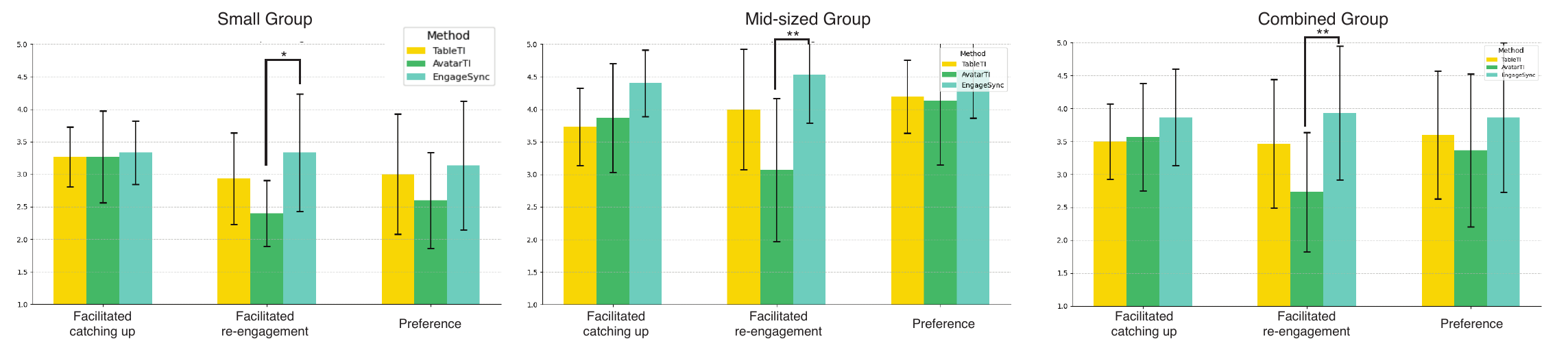}
	\vspace{-1.0em}
	\caption{Statistical results of utility-related questions. Results showed that participants perceived EngageSync to facilitate re-engagement compared to AvatarTI. No statistical differences were observed for perceived catch-up facilitation and preference.}
	\label{fig:util}
\end{figure*}

\subsubsection{Cognitive Load and Usability}
\hfill \break
\textbf{Cognitive Load :}
The NASA-TLX scores revealed no significant differences in the small group condition across the three interfaces (see Figure~\ref{fig:nasa}). This suggests that in smaller groups, the cognitive effort required to use EngageSync, AvatarTI, and TableTI was comparable.
In the mid-sized group, however, TableTI induced significantly higher cognitive load compared to both EngageSync (\textit{p} < .05) and AvatarTI (\textit{p} < .05), with a moderate-to-large effect size (W = .437). A similar trend was observed across all participants in the combined group, where TableTI (\textit{Mdn} = 22.0) led to significantly higher cognitive load than EngageSync (\textit{Mdn} = 18.0, \textit{p} < .01) and AvatarTI (\textit{Mdn} = 19.0, \textit{p} < .05). Notably, there was no significant difference between the two avatar-fixed panels; AvatarTI and EngageSync (\textit{p} = .19), indicating that the avatar-fixed solutions maintained a relatively low cognitive load in all conditions.

In terms of group size, participants in the mid-sized group reported significantly higher cognitive load when using TableTI compared to the small group (\(U = 50.0, p =0.0084\)). This finding is further supported by participant feedback, as P9 explained, \textit{“It took extra effort to keep track of who was speaking and where they were, especially when there were so many people.”} This suggests that as the group size increases, the cognitive demands of processing full transcripts on the tabletop interface become more pronounced, while EngageSync and AvatarTI remain more manageable.

\vspace{1em}
\textbf{Usability: }
Despite these differences in cognitive load, no significant differences in usability (SUS scores) were observed across the interfaces in either group size (small or mid-sized) or when combining all participants (see Figure~\ref{fig:sus}). This suggests that, even though TableTI imposed higher cognitive load, particularly in mid-sized groups, participants did not report a negative impact on overall usability. There were no significant group size effects for usability scores either (\(U = 103.5, p = .29\)).

This finding indicates that while TableTI requires more cognitive effort, particularly in larger groups, it does not detract from its usability. Participants may still find it helpful for full transcripts, despite the increased effort, while EngageSync and AvatarTI offer similar usability with less cognitive load.

\vspace{-1em}
\subsubsection{Utility Questions} 
We asked participants to rate the utility of each interface on a 7-point Likert scale, focusing on how well the interfaces facilitated catching up and re-engagement after disruptions (see Figure~\ref{fig:util}).

\textbf{Facilitated catching up:} 
Participants were asked how well each interface helped them catch up with the conversation after being disrupted. Across all conditions (small, mid-sized, and combined groups), there were no significant differences in participants' ratings of the interfaces' ability to help them catch up. This suggests that participants felt equally capable of catching up using EngageSync, AvatarTI, and the tabletop interface, regardless of group size or interface type (\(U = 101.6, p = .28\)).

\begin{figure*}[h]	
\centering\includegraphics[width=1.0\textwidth]{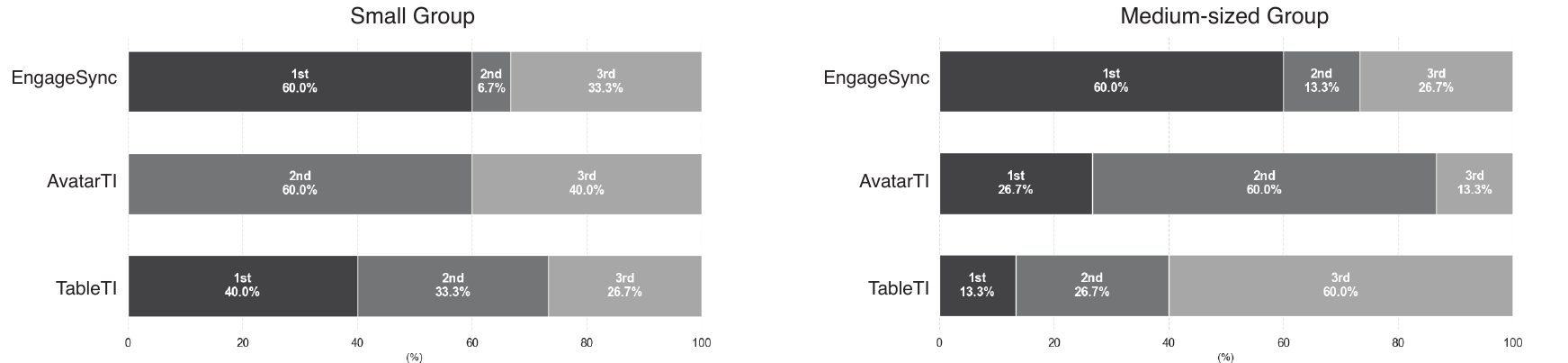}
	\vspace{-2.0em}
	\caption{Ranked Preference Scores across different group conditions. EngageSync was consistently ranked between small and mid-sized group conditions with over half of the participants preferring it. }
	\label{fig:rank}
\end{figure*}

\textbf{Facilitated re-engagement:} 
When evaluating how well each interface helped participants re-engage after dropping out, significant differences were found across group sizes. In the small group condition, both EngageSync and the tabletop baseline were rated as more helpful for re-engagement compared to AvatarTI. Participants found it easier to return to the conversation using EngageSync (\textit{Mdn} = 4.0) and the tabletop interface (\textit{Mdn} = 3.0) than with AvatarTI (\textit{Mdn} = 2.0), although no significant difference was observed between EngageSync and the tabletop baseline.

In the mid-sized group, EngageSync performed notably better than both AvatarTI and the tabletop interface. Participants rated EngageSync significantly higher (\textit{Mdn} = 5.0) for helping them re-engage compared to AvatarTI (\textit{Mdn} = 3.0) and the tabletop baseline (\textit{Mdn} = 4.0). P12 shared, \textit{“I liked how the summary panels stayed up when I rejoined, ... I find it harder to match the names of who talked while I was gone when there are too many people.”} This suggests that the adaptive features of EngageSync were particularly effective in facilitating re-engagement in larger group settings, likely due to its ability to present summarized content after periods of disengagement.

Across all participants, EngageSync (\textit{Mdn} = 4.0) was rated as the most helpful interface for re-engagement, outperforming both AvatarTI (\textit{Mdn} = 3.0) and the tabletop interface (\textit{Mdn} = 4.0). These results indicate a clear preference for the context-aware interface, which provided more effective support for rejoining the conversation after disruptions. Additionally, participants in the mid-sized group rated EngageSync as significantly more helpful compared to those in the small group, suggesting that the interface's adaptive features are even more valuable in larger group settings.

\textbf{Ranked Preference :}
Participants ranked their preferences for the three interfaces, and results showed a clear favoring of EngageSync. In the small group condition, 7 participants ranked EngageSync as their first choice, closely followed by 6 participants preferring TableTI, while only 2 chose AvatarTI. In the mid-sized group, preference for EngageSync increased, with 10 out of 15 participants ranking it first, while AvatarTI saw a slight rise in preference (3 participants), and TableTI fell to 2 participants ranking it first, more frequently receiving third-place rankings.

Across all participants, EngageSync was the most preferred interface (17 out of 30), followed by TableTI (8 out of 30) and AvatarTI (5 out of 30). Interestingly, while AvatarTI's preference increased in mid-sized groups, TableTI remained consistently less favored, suggesting that the on-demand nature of EngageSync was better suited for both group sizes, with AvatarTI becoming slightly more acceptable in larger groups.

\section{Discussion}

\subsection{The Re-engagement and Social Presence Trade-off}
One of the most significant findings in our study was the trade-off between maintaining social engagement and efficiently keeping up with the conversation in immersive meetings, particularly when comparing the avatar-fixed (AvatarTI, EngageSync) and table-fixed (TableTI) interfaces. Some participants preferred TableTI for its familiarity. P3 noted, \textit{“I'm used to seeing everything in one place, it feels natural to me, like how I use Zoom."} However, this convenience came at the cost of social presence. Gaze tracking data revealed that participants using TableTI spent significantly more time focusing on the centralized transcription panel instead of other avatars, leaving them feeling disconnected, supporting \textbf{H2}. As P7 reflected, \textit{"I liked having everything in front of me, but I realized I was just staring at the text... I didn't feel like I was in a meeting. For a second, I forgot I was in VR as well."} P11 echoed this detachment, stating, \textit{“I caught up with the conversation, but I didn’t really feel like I was part of it.”}

In contrast, AvatarTI mitigated some of these issues by aligning users’ gaze more closely with the group, thus enhancing social presence. Nevertheless, it lacked the ability to provide context-sensitive summaries, an important feature for users re-entering the meeting after disruptions. EngageSync addressed this gap by dynamically adapting to users’ engagement state, generating summaries that helped participants efficiently catch up on missed content without sacrificing their connection to the group. This finding supports \textbf{H1}, showing that EngageSync not only improved users' ability to keep up with conversations after disruptions but also preserved social presence more effectively than both TableTI and AvatarTI.

Ultimately, the trade-off between re-engagement and social presence highlights the need for more nuanced transcription interfaces in immersive settings. While traditional setups like TableTI offer a familiar, one-stop display of conversation data, they risk isolating users from the social dynamics of the meeting. By contrast, spatially distributed approaches such as EngageSync strike a more delicate balance, enabling re-engagement without sacrificing immersion. Going forward, interface designs should prioritize flexibility and adaptability so that users can preserve social connections while effectively managing the flow of meeting information.

\subsection{Remembering the Less Talkative Person’s Comments}
Participants often struggled to recall comments from quieter individuals, especially in mid-sized groups where conversation dynamics become more complex. This challenge surfaced most clearly with TableTI, as participants frequently shifted their focus from the group to the transcription panel. In line with prior work on meeting inclusivity~\cite{hosseinkashi2024meeting}, this centralized design can inadvertently marginalize lower-frequency speakers by drawing attention away from them. P5 remarked, \textit{“It was hard to keep track of what everyone was saying, especially the person who didn’t speak much,”} highlighting how the table-fixed interface may dilute awareness of quieter contributions.

In contrast, avatar-fixed interfaces like AvatarTI and EngageSync anchored the text directly to each speaker’s location, reinforcing both memory and social presence. P7 noted, \textit{“Having the panel attached to the avatar really helped me remember who said what. I could recall both their position and their comments more easily,”} illustrating how spatial cues help participants align spoken contributions with specific individuals. This spatialized design ensures that even quieter voices receive visible representation, particularly helpful in larger, fast-paced conversations.

While further validation is needed, these initial findings suggest that spatialized transcription interfaces hold promise for promoting equity in meetings by distributing attention more evenly across all participants. By anchoring transcriptions to avatars, EngageSync could help foster a more inclusive environment, where every voice is remembered and valued, contributing to a more balanced and collaborative meeting dynamic. Future research could explore the long-term impact of such interfaces on meeting inclusivity, particularly in more diverse and larger group settings.

\subsection{Re-engagement Strategies}
Participants adopted varying strategies to catch up on missed content depending on the interface. With EngageSync, two main approaches emerged: most participants read all summary panels before fully rejoining the live discussion, while others preferred to listen to the current speaker first, then consult the summaries. As P8 explained, \textit{“It was useful as I listened and tried to match that with what they had said while I was gone.”} Some who chose to read everything up front felt rushed. P3 noted, \textit{“Since I knew I had to read everything to return to the default mode, I tried to comprehend the summaries as fast as I could.”} This urgency highlights the need for smoother transitions between summary and live modes.

Meanwhile, TableTI users often scrolled rapidly through the text panels to spot key points before resuming active listening. Yet in mid-sized groups, several participants (P18, P23, P26) reported feeling overwhelmed by the volume of text and admitted to ``giving up” on catching up entirely. They cited difficulties in keeping pace with ongoing conversations while also parsing large amounts of backlog.

For AvatarTI, the majority of participants (23 out of 30) opted to listen to the current speaker and attempt to fill in the gaps, rather than relying solely on summary panels. Although some still tried to catch up through the summaries, they found it less effective than simply re-engaging with the live discussion.

Overall, these patterns reveal that re-engagement is not a one-size-fits-all process. Future designs might adapt to diverse user preferences, whether that means gradually introducing missed content, offering quick summaries, or allowing a smoother shift from summary review back to live interaction.

\subsection{Group Size Effects on Interface Effectiveness}
Our findings provide strong evidence that avatar-fixed transcription tools become increasingly beneficial in larger groups, thereby supporting \textbf{H4}. Participants in the mid-sized group expressed a clear preference for avatar-fixed panels—especially EngageSync—because they needed to manage more speakers and multiple conversation threads. This heightened complexity made the advantages of spatially anchored transcription more apparent.

In the mid-sized group, EngageSync significantly outperformed TableTI on attention allocation and re-engagement time. By attaching panels to each avatar, participants found it easier to stay immersed in the conversation flow. As P7 noted, \textit{“With the avatar-fixed panel, I felt more in sync with the conversation ... it was easier to follow who was saying what.”} This alignment of text with speakers not only helped participants track the conversation more intuitively but also reduced the mental effort of identifying each speaker’s contributions.

Cognitive load was notably higher with TableTI in mid-sized groups, reflecting the burden of mentally mapping multiple voices to a single, centralized transcript. In contrast, EngageSync’s context-sensitive adaptation further aided users in quickly catching up after dropouts, a critical need in more complex meetings. Participants valued how summary panels appeared directly where they were needed, allowing them to resume the conversation faster.

In smaller group settings, although participants still favored avatar-fixed interfaces over TableTI, the differences in attention allocation and cognitive load were less pronounced. With fewer speakers, it was simpler to identify who was talking, even with a table-fixed panel. Nonetheless, our results show that in larger groups, where conversation dynamics are more involved, avatar-fixed solutions—particularly EngageSync—are essential for maintaining social presence and supporting smooth re-engagement.
\section{Limitations and Future Work}

One limitation of this study is the fixed placement of text panels above avatars. While this placement was chosen to minimize interference, alternative configurations (e.g., left, right, or below avatars) may affect attention, cognitive load, and social presence differently. Future work could explore how these variations impact user experience in various meeting contexts. 

Another limitation relates to our experimental design choices. The study used a four-minute dropout to simulate real-world interruptions, which our formative study participants confirmed as representative of common meeting interruptions like taking phone calls or brief absences. However, different types of disengagements (e.g., shorter or longer absences) could produce varying effects on re-engagement. 
Additionally, our use of pre-recorded conversations meant participants were passive listeners, which may have affected their recall and social presence compared to live meetings. For example, in a live setting, users' re-engagement strategies might differ from those in a pre-recorded setting, particularly since ignoring a current speaker to read a listener's previous contributions would have more noticeable social implications. Future studies should examine different interruption scenarios~\cite{son2023okay} and investigate these interfaces in real-time interactive settings to assess their impact during active participation.

The presentation and structure of information also present opportunities for improvement. Our basic presentation of missed conversations aimed to prevent clutter, but future research could enhance both the order and presentation of missed content. 
This could involve adapting spatiotemporal visualizations~\cite{andrienko2003exploratory} 
 or exploring hybrid interfaces that allow adaptive UI placement from world- to screen-fixed, as suggested by previous work~\cite{pei2024ui}. Such improvements could provide users with more flexible and intuitive ways to access missed information.

Finally, the technical accuracy of our system remains an area for future development. The accuracy of the ASR-to-LLM-generated summaries continues to be a challenge. While this study focused primarily on how to display the summaries, future work should address the fundamental issue of improving summary generation to enhance the overall reliability and effectiveness of the system.

\section{Conclusion}
In this paper, we introduced EngageSync, a context-aware avatar-fixed panel, and demonstrated its effectiveness in enhancing both social presence and information recall in immersive meetings. Our study showed that EngageSync outperformed traditional Tabletop and always-on Avatar-fixed panels, particularly in mid-sized groups, where maintaining engagement and catching up after disruptions posed greater challenges. These results reinforce the findings from our formative study, which identified the need for context-sensitive transcription methods in VR environments.

The simplicity of adapting transcription panels based on user engagement is key to EngageSync's effectiveness. Our findings suggest that by providing live transcriptions and summaries when necessary, users are better equipped to re-engage with ongoing discussions without sacrificing social presence. This adaptive approach can be seamlessly integrated into current VR meeting platforms, providing a user-friendly solution to a persistent challenge in immersive meetings.

Future research could examine how context-aware interfaces like EngageSync can be further refined or expanded. For example, combining avatar-fixed panels with more flexible, user-controlled interactions, or exploring the role of different panel placements, could yield even more efficient designs. We hope this study inspires further innovation in adaptive VR interfaces, promoting more effective and natural interactions in immersive environments.

\bibliographystyle{ACM-Reference-Format}
\bibliography{_showYourOp}

\end{document}